\documentclass[twocolumn]{aastex7}

\usepackage{amssymb, amsmath}
\usepackage{graphicx}
\usepackage{color}
\usepackage{ulem}
\usepackage{appendix}
\usepackage{hyperref}
\usepackage[stable]{footmisc}
\usepackage{relsize}

\graphicspath{{./}{figures/}{lognormal_plots/}{norv/}}
\begin{document}

\title{Multi-band Spectral and Astrometric Characterization of the HIP 99770 b Planet with SCExAO/CHARIS and Gaia\footnote{Based in part on data collected at Subaru Telescope, which is operated by the National Astronomical Observatory of Japan.}}

\author{Danielle Bovie}
\affiliation{Department of Physics and Astronomy, University of Texas-San Antonio, San Antonio, TX, USA}
\email[show]{danielle.bovie@my.utsa.edu}
\author{Thayne Currie}
\affiliation{Department of Physics and Astronomy, University of Texas-San Antonio, San Antonio, TX, USA}
\email[]{thayne.currie@utsa.edu}
\affiliation{Subaru Telescope, National Astronomical Observatory of Japan, 
650 North A`oh$\bar{o}$k$\bar{u}$ Place, Hilo, HI  96720, USA}
\author{Mona El Morsy}
\affiliation{Department of Physics and Astronomy, University of Texas-San Antonio, San Antonio, TX, USA}
\email[]{mona.elmorsy@utsa.edu}
\author{Brianna Lacy}
\affiliation{Department of Astronomy and Astrophysics, University of California-Santa Cruz, Santa Cruz, CA USA}
\email[]{blacy@ucsc.edu}
\author{Masayuki Kuzuhara}
\affiliation{Astrobiology Center of NINS, 2-21-1, Osawa, Mitaka, Tokyo, 181-8588, Japan}
\affiliation{National Astronomical Observatory of Japan, 2-21-2, Osawa, Mitaka, Tokyo 181-8588, Japan}
\email[]{m.kuzuhara@nao.ac.jp}
\author{Jeffrey Chilcote}
\affiliation{Department of Physics, University of Notre Dame, South Bend, IN, USA}
\email[]{jchilcot@nd.edu}
\author{Taylor L. Tobin}
\affiliation{Department of Astronomy, University of Michigan, Ann Arbor, MI, USA}
\email[]{tltobin@umich.edu}
\author{Olivier Guyon}
\affiliation{Subaru Telescope, National Astronomical Observatory of Japan, 
650 North A`oh$\bar{o}$k$\bar{u}$ Place, Hilo, HI  96720, USA}
\affil{Astrobiology Center of NINS, 2-21-1, Osawa, Mitaka, Tokyo, 181-8588, Japan}
\affil{Steward Observatory, The University of Arizona, Tucson, AZ 85721, USA}
\affil{College of Optical Sciences, University of Arizona, Tucson, AZ 85721, USA}
\email[]{guyon@naoj.org}
\author{Tyler D. Groff}
\affiliation{NASA-Goddard Space Flight Center, Greenbelt, MD, USA}
\email[]{tyler.d.groff@nasa.gov}
\author{Julien Lozi}
\affiliation{Subaru Telescope, National Astronomical Observatory of Japan, 
650 North A`oh$\bar{o}$k$\bar{u}$ Place, Hilo, HI  96720, USA}
\email[]{lozi@naoj.org}
\author{Sebastien Vievard}
\affiliation{Subaru Telescope, National Astronomical Observatory of Japan, 
650 North A`oh$\bar{o}$k$\bar{u}$ Place, Hilo, HI  96720, USA}
\email[]{vievard@naoj.org}
\author{Vincent Deo}
\affiliation{Subaru Telescope, National Astronomical Observatory of Japan, 
650 North A`oh$\bar{o}$k$\bar{u}$ Place, Hilo, HI  96720, USA}
\affiliation{Optical Sharpeners SAS, Manosque, France}
\email[]{vdeo@naoj.org}
\author{Frantz Martinache}
\affiliation{Universit\'{e} C\^{o}te d'Azur, Observatoire de la C\^{o}te d'Azur, CNRS, Laboratoire Lagrange, France}
\email[]{Frantz.Martinache@oca.eu}
\author{Yiting Li}
\affiliation{Department of Physics, University of California, Santa Barbara, Santa Barbara, California, USA}
\email[]{lyiting@umich.edu}
\author{Motohide Tamura}
\affil{Astrobiology Center of NINS, 2-21-1, Osawa, Mitaka, Tokyo, 181-8588, Japan}
\affiliation{National Astronomical Observatory of Japan, 2-21-2, Osawa, Mitaka, Tokyo 181-8588, Japan}
\affiliation{Department of Astronomy, Graduate School of Science, The University of Tokyo, 7-3-1, Hongo, Bunkyo-ku, Tokyo, 113-0033, Japan}
\email[]{motohide.tamura@nao.ac.jp}

\begin{abstract}

We present and analyze follow-up, higher resolution ($R$ $\sim$ 70) $H$ and $K$ band integral field spectroscopy of the superjovian exoplanet HIP 99770 b with SCExAO/CHARIS. Our new data recover the companion at a high signal-to-noise ratio in both bandpasses and more than double the astrometric baseline for its orbital motion. Jointly modeling HIP 99770 b's position and the star's astrometry from \textit{Hipparcos} and \textit{Gaia} yields orbital parameters consistent with those from the discovery paper, albeit with smaller errors, and a slight preference for a smaller semimajor axis ($\sim$15.7--15.8 au)and a larger eccentricity ($\sim$0.28--0.29), disfavoring a circular orbit. We revise its dynamical mass slightly downwards to 15.0$_{-4.4}^{+4.5}$ $M_{\rm Jup}$ for a flat prior and 13.1$_{-5.2}^{+4.8}$ $M_{\rm Jup}$ for a more standard log-uniform mass prior, where the inclusion of its relative radial-velocity measurement is primarily responsible for these changes.  We find consistent results for HIP 99770 b's dynamical mass including recent VLTI/GRAVITY astrometry, albeit with a slightly smaller, better constrained eccentricity of $e$ $\sim$ 0.22$^{+0.10}_{-0.13}$.  HIP 99770 b is a $\sim$ 1300 K  object at the L/T transition with a gravity intermediate between that of the HR 8799 planets and older, more massive field brown dwarfs  with similar temperatures but with hints of equilibrium chemistry. 

HIP 99770 b is particularly well suited for spectroscopic follow up with Roman CGI during the technology demonstration phase at 730 nm to further constrain its metallicity and chemistry; JWST thermal infrared observations could likewise explore the planet's carbon chemistry, metallicity, and clouds.

\end{abstract}


\section{Introduction}

Of the more than 5,000 exoplanets discovered around nearby stars, $\sim$25 have been directly imaged, primarily using facility and now dedicated \textit{extreme} adaptive optics systems \citep[e.g.][]{Marois2008a,Lagrange2010,Currie2023a}. The demographics of these discoveries shed light on the primary modes of gas giant planet formation \citep{Nielsen2019}. Follow-up characterization has led to breakthroughs in the understanding of clouds, chemistry, and gravity-sensitive features of young exoplanets and how they differ from those of field brown dwarfs \citep{Currie2011,Barman2011a,Barman2015,Konopacky2013}. 

In the past few years, many direct imaging searches have focused on targets that show dynamical evidence of a companion, usually from precision \textit{Gaia} and \textit{Hipparcos} astrometry \citep[e.g.][]{Kuzuhara2022,Currie2023b, Franson2023b}. The $\sim$24 year baseline between the \textit{Hipparcos} and \textit{Gaia}-DR3 instantaneous proper motion measurements can reveal a nearby bright star's astrometric acceleration induced by a substellar companion within 1\arcsec{} (or $\le$50 au for a star at 50 pc). These detection regions for astrometry are well matched to the angular separations over which leading extreme AO systems can image planets. Compared to the ``blind"/unbiased approach typical of earlier campaigns, where targets were selected based on system age and distance, this approach appears to be more successful in producing a higher discovery rate of planets and brown dwarfs \citep{ElMorsy2024b}.

Furthermore, planets and brown dwarfs detected from both direct imaging and astrometry can be characterized in greater depth. Jointly modeling the star's astrometry with the companion's relative astrometry from the imaging data directly constrain the companion's dynamical mass, avoiding the fundamental uncertainties associated with luminosity evolution models used when only imaging data are available \citep{Brandt2019}. Dynamical modeling of astrometry and imaging data also better constrains orbits. Coupling information on the companion's mass and orbit, separate constraints on the system's age, and the constraints on atmospheres provided by direct imaging data (e.g. with low-resolution integral field spectrographs) helps probe how the luminosity and atmospheric properties of substellar objects evolve \citep[e.g.][]{GBrandt2021,Franson2023,ElMorsy2024b}.

HIP 99770 b is the first superjovian planet detected by direct imaging and astrometry and an example of the power for this combined approach to more holistically characterize companions \citep{Currie2023b}. The primary star (HIP 99770 A) has a spectral type of A5V, an asteroseismology-determined age of $\approx$115--414 $Myr$, and is surrounded by a cold debris disk. Data from the Hipparcos-Gaia Catalogue of Accelerations \citep[HGCA;][]{Brandt2021a} reveal potential evidence for a companion from the primary's proper motion anomaly \citep[$\chi^2\sim$ 7.23; ][]{Currie2023b}. SCExAO/CHARIS \citep{Jovanovic2015,Groff2016} and Keck/NIRC2 high-contrast imaging data then resolved HIP 99770 b at a projected separation of $\approx$0\farcs{}43--0\farcs{}45 over six data sets between 2020 and 2021.

Dynamical modeling of HIP 99770 b found a best-estimated mass of $\sim$13.9--16.1 $M_{\rm Jup}$, a semimajor axis of $\approx$17 au, a low inclination ($\sim$30$^{o}$ from face on), and a low eccentricity ($e$ $\sim$ 0.25) \citep{Currie2023b}. Atmospheric modeling suggests that HIP 99770 b lies near the L/T transition with cloud thicknesses intermediate between younger, slightly lower-mass exoplanets around HR 8799 and older, higher-mass field brown dwarfs. Follow-up high-resolution coronagraphic spectroscopy with Keck/KPIC resolved H$_{\rm 2}$O and CO in HIP 99770 b's atmosphere, yielded a carbon-to-oxygen ratio of C/O $\sim$ 0.55--0.64, and metallicity consistent with solar \citep{Zhang2024}. Compared to other young superjovian companions, HIP 99770 b may exhibit a slightly slow rotation rate.

Follow-up astrometric modeling and integral field spectroscopy (IFS) could improve our estimates for HIP 99770 b's dynamics and atmosphere. Over the 15-month astrometric baseline modeled in \citet{Currie2023b}, HIP 99770 b moved $\sim$7$^{o}$ clockwise with a small net decrease in its projected separation. As suggested by Figures 2 and especially S9 in the discovery paper, a doubling of HIP 99770 b's astrometric baseline could further refine the orbital properties and mass, especially when coupled with relative RV measurements from \citet{Zhang2024}. HIP 99770 b currently has very low-resolution ($R$ $\sim$ 20) CHARIS spectra (1.1--2.4 $\mu m$) and NIRC2 photometry primarily capable of probing its temperature and very high spectral resolution data sensitive to narrow molecular lines. Other directly imaged planets and brown dwarfs with dynamical masses like HR 8799 cde and HD 33632 Ab have slightly higher resolution IFS data \citep[e.g.][]{Greenbaum2018,Nasedkin2024,ElMorsy2024}. Similar data for HIP 99770 b will provide a context for the companion's gravity and carbon chemistry, which can be probed by features in $H$ and $K$ band, respectively \citep[e.g.][]{AllersLiu2013,Barman2011a}.

In this paper, we present follow-up higher-resolution ($R$ $\sim$ 70) $H$ and $K$ band data for HIP 99770 b obtained with SCExAO/CHARIS. These data more than double the available astrometric baseline for HIP 99770 b and advance our understanding of HIP 99770 b's atmosphere. We use our results to investigate the calibration of luminosity evolution models using substellar companions with measured dynamical masses. Finally, we forecast the suitability of HIP 99770 b for the \textit{Roman Space Telescope}'s Coronagraphic Instrument, as a part of its technological demonstration and potential follow-on imaging and spectroscopic science observations.

\begin{deluxetable*}{llllllllll}
     \tablewidth{0pt}
    \tablecaption{HIP 99770 Observing Log\label{tab:obslog}}
    \tablehead{\colhead{UT Date} & \colhead{Instrument} &  \colhead{Seeing$^{b}$ (\arcsec{})} &{Filter} & \colhead{$\lambda$ ($\mu m$)$^{a}$} & \colhead{Occulting mask radius}
    & \colhead{$t_{\rm exp}(s)$} & \colhead{$N_{\rm exp}$} & \colhead{$\Delta$PA ($^{o}$)} & SNR}
    \startdata
    20230606 & SCExAO/CHARIS$^{a}$ &  0.6-0.8 & $K$ & 2.01--2.36 & 0\farcs{}198 & 60.48 & 120 & 107.11  & 20.78\\
    20230708 & SCExAO/CHARIS$^{a}$ &  0.5 & $H$ & 1.47--1.79 & 0\farcs{}228 & 41.3 & 183 & 70.918 & 25.64\\
    \enddata
    \tablecomments{
    a) This column refers to the CHARIS IFS wavelength range. b) From the Canada France Hawaii Telescope seeing monitor.    
    }
    \end{deluxetable*}

   \begin{figure*}
    \centering
       \vspace{-0.4in}
    \includegraphics[width=1\textwidth,trim=28mm 15mm 30mm 30mm,clip]{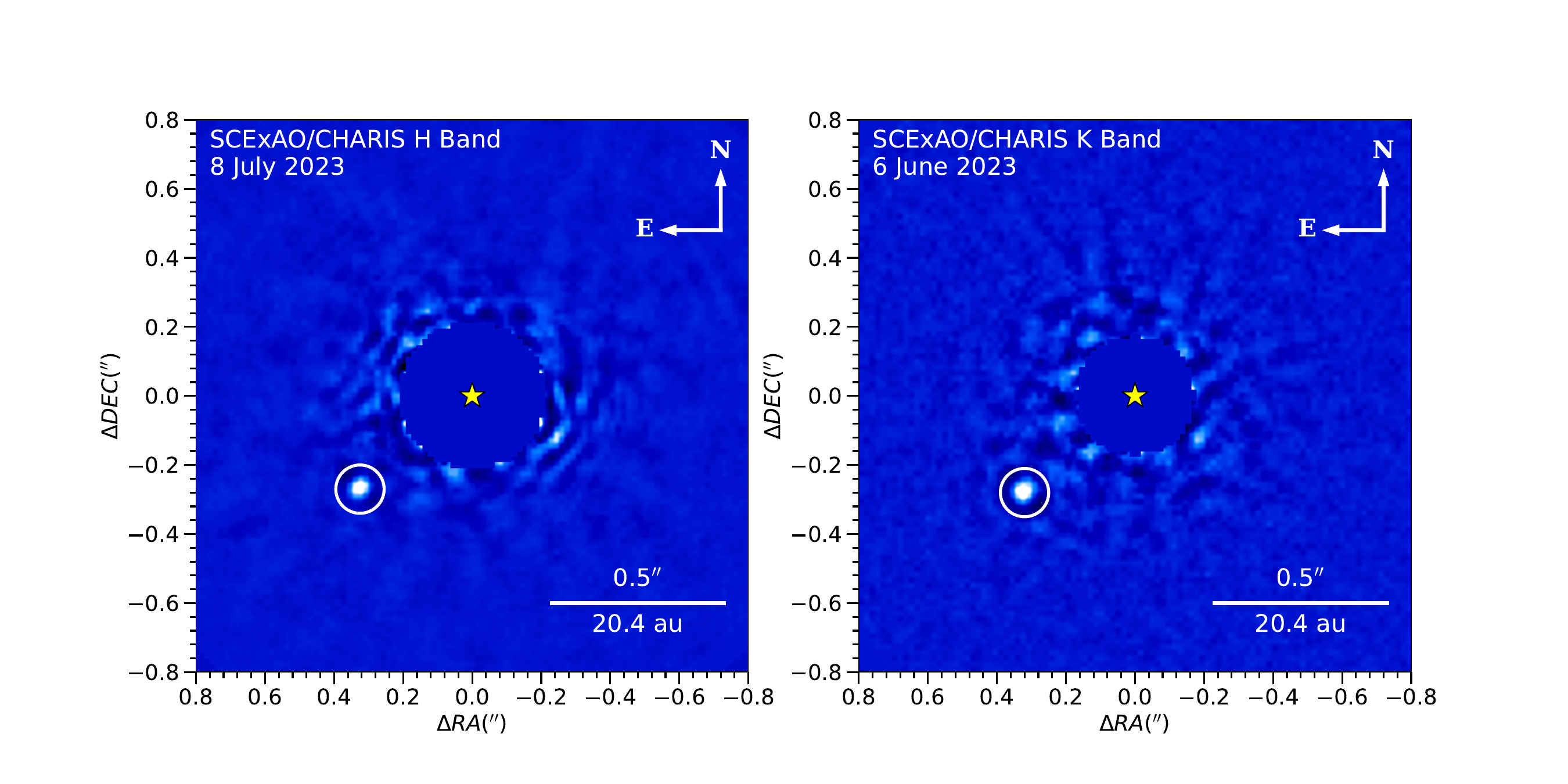}
    \vspace{-0.1in}
    \caption{HIP 99770 b SCExAO/CHARIS detections in the H and K bands on 8 July 2023 and 6 June 2023 respectively. The detected companion is circled and the centroid of the star is marked by a yellow star.}
    \vspace{-0.in}
    \label{fig:images}
\end{figure*}

\begin{deluxetable*}{llllllllll}
     \tablewidth{0pt}
    \tablecaption{HIP 99770 b Astrometry\label{tab:astrom}}
    \tablehead{\colhead{UT Date} & \colhead{E(\arcsec{})} &  \colhead{N(\arcsec{})} & \colhead{Separation(\arcsec{})}
    & \colhead{Position angle($^{o}$)}}
    \startdata
    $20230606$ &  $0.3219 \pm 0.0035$ &  $-0.2764 \pm 0.0034$ & $0.4243\pm 0.0034 $ & $130.66\pm 0.49$\\
    $20230708$ &  $0.3242 \pm 0.0035$ & $-0.2650 \pm 0.0034$ & $0.4187\pm0.0034 $  & $129.27\pm 0.49$\\
    \enddata
    \end{deluxetable*}

\begin{deluxetable}{llll}
     \tablewidth{0pt}
    \tablecaption{HIP 99770 b Spectrum\label{tab:spec}}
    \tablehead{\colhead{Wavelength ($\mu$m)} & \colhead{$F_{\rm \nu}$ (mJy)} &  \colhead{$\sigma$~$F_{\rm \nu}$ (mJy)} & \colhead{SNR}}
    \startdata
1.477 & 0.072 & 0.032 & 2.223 \\ 
1.492 & 0.154 & 0.024 & 7.381 \\ 
1.508 & 0.117 & 0.020 & 5.967 \\ 
1.523 & 0.160 & 0.020 & 8.254 \\ 
1.539 & 0.186 & 0.019 & 9.906 \\ 
1.554 & 0.206 & 0.019 & 11.655 \\ 
1.570 & 0.231 & 0.021 & 12.516 \\ 
1.586 & 0.285 & 0.020 & 15.872 \\ 
1.602 & 0.275 & 0.020 & 15.569 \\ 
1.618 & 0.331 & 0.021 & 19.107 \\ 
1.635 & 0.334 & 0.021 & 18.163 \\ 
1.652 & 0.343 & 0.021 & 21.164 \\ 
1.668 & 0.313 & 0.019 & 17.568 \\ 
1.685 & 0.362 & 0.020 & 23.361 \\ 
1.702 & 0.286 & 0.018 & 17.745 \\ 
1.720 & 0.272 & 0.017 & 17.645 \\ 
1.737 & 0.228 & 0.022 & 11.225 \\ 
1.755 & 0.197 & 0.020 & 10.224 \\ 
1.773 & 0.127 & 0.026 & 6.263 \\ 
1.791 & 0.130 & 0.109 & 1.236 \\ 
\hline
2.015 & 0.253 & 0.077 & 7.589 \\ 
2.036 & 0.249 & 0.021 & 12.581 \\ 
2.056 & 0.304 & 0.026 & 12.604 \\ 
2.077 & 0.317 & 0.024 & 14.726 \\ 
2.098 & 0.357 & 0.026 & 15.100 \\ 
2.119 & 0.372 & 0.025 & 16.587 \\ 
2.141 & 0.379 & 0.028 & 15.072 \\ 
2.163 & 0.410 & 0.033 & 13.389 \\ 
2.184 & 0.407 & 0.032 & 13.713 \\ 
2.207 & 0.360 & 0.029 & 13.218 \\ 
2.229 & 0.332 & 0.033 & 10.849 \\ 
2.252 & 0.308 & 0.037 & 8.833 \\ 
2.274 & 0.333 & 0.042 & 8.475 \\ 
2.297 & 0.281 & 0.043 & 6.918 \\ 
2.321 & 0.204 & 0.043 & 4.909 \\ 
2.344 & 0.207 & 0.041 & 5.307 \\ 
2.368 & 0.204 & 0.061 & 3.930 \\ \hline
\enddata
\tablecomments{A horizontal divider is placed between H band spectral measurements (top) and K band measurements (bottom).}

\end{deluxetable}

\begin{figure}
\centering
   \includegraphics[width=0.5\textwidth,trim=5mm 0mm 12mm 12mm,clip]{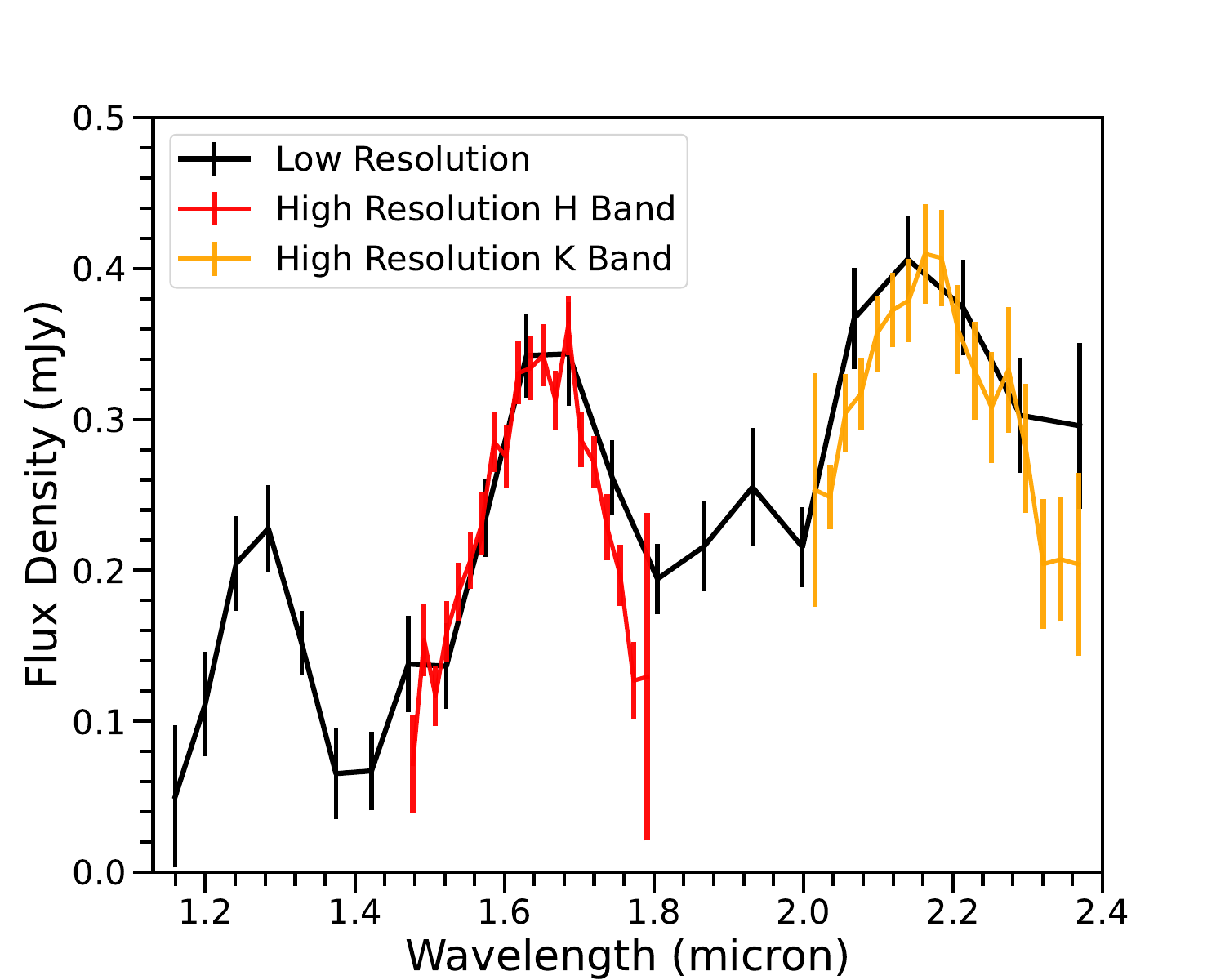}
   \caption{HIP 99770 b H band spectra in red from 8 July 2023 and K band spectra in orange from 6 June 2023 plotted on top of the low resolution spectra from the original discovery in black from 17 October 2021 and published in \citet{Currie2023b}.}
   \label{fig:spectra} 
\end{figure}

\section{Data}

\subsection{Observations}

We observed HIP 99770 b with SCExAO/CHARIS in the H (1.47–1.79$\mu$m) and K (2.01–2.36 $\mu$m) passbands on 8 July 2023 and 5 June 2023 (Table \ref{tab:obslog}).
Conditions were photometric on both nights. The K band data were obtained under average to slightly below-average seeing conditions ($\theta_{\rm V}$ = 0\farcs{}6-0\farcs{}8) and $\sim$10 mph winds; for H band, the seeing was better ($\theta_{\rm V}$ = 0\farcs{}5) but with 15-20 mph winds. We acquired all data in pupil tracking such that the sky rotates in the frame of the detector, enabling \textit{angular differential imaging}  \citep[ADI][]{Marois2006}.

Our H band data consist of 241 exposures of 41.3 seconds each, with a total parallactic angle rotation of 70.918$^{o}$. In K band, our sequence was longer (195 exposures of 60.48 seconds) and covered a larger parallactic angle rotation (107.11$^{o}$). For all data, we used a classic Lyot coronagraph to suppress the stellar halo and generated satellite spots for spectrophotometric and astrometric calibration by modulating the SCExAO deformable mirror \citep[25 nm modulation amplitude][]{Jovanovic2015-astrogrids} . For the K band we also took 5 sky frames prior to the science observations.

\subsection{Data Reduction}
Our data reduction approach largely followed that of previous SCExAO/CHARIS observations of accelerating stars \citep{Currie2020a,Kuzuhara2022,Currie2023b,Tobin2024}. We extracted the CHARIS data cubes using the reduction pipeline from \citet{Brandt2017}\footnote{https://github.com/PrincetonUniversity/charis-dep}. We used the recommended default settings and for calibration took dark frames and narrowband flats on the same nights, immediately prior to observations. Subsequent data reduction steps utilize the CHARIS Data Processing Pipeline \citep{Currie2020a}. For the K band data, we first removed sky thermal emission by subtracting separate sky cubes. We registered each cube to a common center using a two-degree polynomial fit to the satellite spot positions.  

Inspection of the registered cubes reveals some variation in image quality, especially for K band. To remove cubes with degraded raw contrasts, we computed the average halo brightness between 0\farcs{}2 and 0\farcs{}4 from the star relative to that of the satellite spots in the bluest channel for each filter. We retained cubes in H (K) band with a spot-to-halo brightness ratio $\geq$ 20 (10). We calibrated the spectrophotometry of the remaining cubes using a Kurucz stellar atmosphere model \citep{Castelli2003} appropriate for an A5V star. Finally, we subtracted a moving box median-filter from each slice of each cube with lengths of $\approx$8 (4) $\lambda/D$ on a side in H (K) band.

To subtract the stellar halo, we used the Adaptive, Locally-Optimized Combination of Images algorithm in combination with ADI \citep[A-LOCI][]{Currie2012,Currie2015}. A-LOCI leverages on free parameters such as the rotation gap as a fraction of the planet full-width at half-maximum ($\delta$), area over which a reference PSF is computed ($N_{\rm A}$), singular value decomposition cutoff ($SVD_{\rm lim}$), and the radial width of the region over which PSF subtraction is applied \citep[see also][]{Lafreniere2007,Marois2010b,Currie2012}. It also controls the number of frames used to build the reference PSF ($n_{\rm ref}$) via a correlation-based frame selection and the option to mask the subtraction region when computing the reference PSF.  

We explored the A-LOCI parameter space to identify combinations that yielded high SNR detections without severe signal loss (see below). HIP 99770 b is detected at a high significance ($>$10-15$\sigma$) in both $H$ and $K$ over a wide range of A-LOCI parameter space. For H band we settled on a wide rotation gap ($\delta$=0.85), used an optimization area of 80 PSF footprints, a singular value decomposition cutoff of SVD$_{lim}=10^{-6}$, and aggressively truncated the reference library ($n_{\rm ref}$ = 60). For K band, we also used a singular value decomposition cutoff of SVD$_{lim}= 10^{-6}$ but our rotation gap and optimization area were smaller ($\delta$=0.5, $N_{\rm A}$ = 50), while we retained more reference frames ($n_{\rm ref}$ = 80), and employed pixel masking.



\subsection{Detections, Spectra, and Astrometry}
Figure \ref{fig:images} shows the final PSF subtracted\footnote{We perform PSF subtraction in combination with ADI only, not ADI+spectral differential imaging (SDI). Utilizing ASDI would result in a higher SNR detection. However, given the $\sim$ narrow bandpass in the CHARIS $H$ and $K$ spectroscopic modes combined with HIP 99770 b's small angular separation, ASDI would result in signal loss that may be more difficult to model.} and wavelength-collapsed images. We decisively detect HIP 99770 b in both $H$ band (left panel) and $K$ band (right panel). Assuming the standard approach for estimating the companion signal-to-noise ratio (SNR) and correcting for finite sample sizes\footnote{We replace each pixel by its sum within a FWHM-sized aperture and compute the robust standard deviation of the (convolved) pixels as a function of angular separation. The initial SNR for HIP 99770 b is the signal at the planet’s location divided by noise at its angular separation. We then determine a final SNR estimate by accounting for small-sample statistics presented in Mawet et al. (2014), such that our SNR values imply a false-alarm probability equivalent to gaussian noise.} \citep{Currie2011,Mawet2014}, the PSF subtracted, wavelength-collapsed $H$ and $K$ band images reveal HIP 99770 b at SNR = 25.64 and 20.78, respectively (see Table \ref{tab:obslog}).

We employed forward-modeling to estimate and correct for signal loss from the planet and astrometric bias due to PSF subtraction following \citet{Currie2018}. The average throughput per channel determined from forward-modeling is roughly $\sim$ $60\%$ for H and $\sim$ $45\%$ for K. The astrometric biasing is a minimal $\sim$0.1--1.2 mas.

Following similar steps from \citet{Kuzuhara2022}, we empirically estimate the uncertainty in HIP 99770 b's position by injecting scaled forward-modeled PSFs at HIP 99770 b's projected separation but at a range of position angles, comparing the inputted and recovered centroid positions. Prior to transforming to polar coordinates, we also adopt a centroiding error of 0.25 pixels in each Cartesian coordinate based on previous experience with the repeatability of astrometric measurements and the scatter of satellite spot-estimated centroids with those drawn from unsaturated PSFs. Typically, our forward-model indicates a jitter of $\sim$2 mas in radial separation and 0.04$^{o}$ in position angle, consistent with the high signal-to-noise ratio of our detections.
For a final astrometric uncertainty, also include the error in the pixel scale (0.05 mas), north position angle (0.27$^{o}$), algorithm biasing (which is negligible), and differences in the H and K band measurements.

Table \ref{tab:astrom} lists the new relative astrometry obtained during the follow-up of HIP 99770 b. HIP 99770 b’s position advanced clockwise from the latest position presented in the object's discovery paper \citep{Currie2023b}. For the two epochs we observed, its average position relative to the primary star was at a position angle (PA) of $\sim$ 129.3--130.7$^{o}$, nearly 10 degrees from its October 2021 position. Its angular separation is also smaller by $\approx$20 mas. The position change between the two epochs we observed (June and July 2023) was also consistent with these trends of clockwise orbital motion and a small inward radial movement.

Table \ref{tab:spec} shows the spectra's flux density, error in flux density, and SNR for each wavelength channel. Our estimate of the spectroscopic uncertainty considers both the intrinsic SNR of the detection and spectrophotometric calibration uncertainty ($\approx 5\%$ per channel). Except for the tellurics-contaminated channels near the extrema of the wavelength ranges for each bandpass, HIP 99770 b is visible at SNR $\gtrsim$ 5--10 per channel over all channels, peaking at SNR $\sim$ 23 at 1.685 $\mu m$.  The spectroscopic uncertainty, which considers the intrinsic SNR of the detection and as well as spectrophotometric calibration uncertainties, is characteristically less than 10--15\% of the measured flux density in each channels.

Figure \ref{fig:spectra} compares our H (red) and K (orange) band spectra to the low-resolution broadband spectra (black) presented in \citet{Currie2023b}. Our high-resolution spectra generally agree with predictions extrapolated from previous measurements within 1-$\sigma$, with the main exceptions occurring in the reddest H and K-band channels, where both our new detections and especially the earlier low-resolution detections were relatively weaker. Integrated over the standard Mauna Kea Observatories $H$ and $K$ passbands, our $H$ and $K$ band spectra yield $m_{\rm H}$ =  16.56 $\pm$ 0.09 and $m_{K_{\rm s}}$ = 15.74 $\pm$  0.10, consistent within errors with the values published in the discovery paper ( $m_{\rm H}$ =  16.51 $\pm$ 0.11 and $m_{K_{\rm s}}$ = 15.66 $\pm$ 0.09).

\begin{deluxetable*}{lccr}
\tablecaption{MCMC Orbit Fitting Priors and Results-- Flat Prior\label{tab:mcmc_results}}
\tablewidth{0pt}
  \tablehead{
    Parameter &
    16/50/84\% quantiles &
    95\% confidence interval &
    Prior}
  \startdata
\multicolumn{4}{c}{Fitted Parameters} \\ \hline
RV jitter (m/s)  &       ${0.08}_{-0.08}^{+50}$ & (0.0, 621) & log-uniform \\
$M_{\rm pri}$ ($M_\odot$)   &      ${1.76}_{-0.15}^{+0.18}$ & (1.48, 2.13) & Gaussian, $1.8 \pm 0.2$ \\
$M_{\rm sec}$ ($M_{\rm Jup}$)  &       ${15}_{-4.4}^{+4.5}$ & (6.6, 25.4) & flat\\
Semimajor axis $a$ (AU)    &     ${15.8}_{-1.0}^{+3.1}$ & (14.1, 24.3) & $1/a$  (log-uniform) \\
$\sqrt{e} \sin \omega$\tablenotemark{\rm *}    &     ${-0.20}_{-0.41}^{+0.44}$ & (-0.67, 0.39) & uniform \\
$\sqrt{e} \cos \omega$\tablenotemark{\rm *}  &       ${0.22}_{-0.40}^{+0.23}$ & (-0.39, 0.54) & uniform  \\
Inclination ($^\circ$)    &     ${151}_{-11}^{+8.8}$   & (131, 165) & $\sin i$ (geometric) \\
PA of the ascending node $\Omega$ ($^\circ$)  &     ${279}_{-269}^{+72}$ & (1, 359) & uniform \\
Mean longitude at 2010.0 ($^\circ$) &        ${120}_{-68}^{+39}$ & (30, 186) & uniform \\
Parallax (mas)   &      ${24.54}_{-0.07}^{+0.07}$ & (24.41, 24.68) & Gaussian, $24.54 \pm 0.07$ \\
\hline
\multicolumn{4}{c}{Derived Parameters} \\ \hline
Period (yrs)     &    ${47}_{-4.4}^{+14}$ & (39, 90) \\
Argument of periastron $\omega$ ($^\circ$)   &      ${250}_{-218}^{+68}$ & (6, 354) \\
Eccentricity $e$    &    ${0.29}_{-0.15}^{+0.12}$ & (0.02, 0.47) \\
Semimajor axis (mas)  &       ${388}_{-25}^{+75}$ & (347, 596) \\
Periastron time $T_0$ (JD)    &     ${2465222}_{-780}^{+3582}$ & (2463666, 2478240) \\
Mass ratio   &      ${0.008}_{-0.002}^{+0.003}$ & (0.004, 0.014)
\enddata
\label{tab:mcmc_results}
\end{deluxetable*}

\begin{deluxetable*}{lccr}
\tablecaption{MCMC Orbit Fitting Priors and Results-- Log-uniform Prior\label{tab:mcmc_results-lognorm}}
\tablewidth{0pt}
  \tablehead{
    Parameter &
    16/50/84\% quantiles &
    95\% confidence interval &
    Prior}
  \startdata
\multicolumn{4}{c}{Fitted Parameters} \\ \hline
RV jitter (m/s)  &       ${0.11}_{-0.11}^{+54}$ & (0.0, 646) & log-uniform \\
$M_{\rm pri}$ ($M_\odot$)   &      ${1.75}_{-0.14}^{+0.17}$ & (1.48, 2.11) & Gaussian, $1.8 \pm 0.2$ \\
$M_{\rm sec}$ ($M_{\rm Jup}$)  &        ${13.1}_{-5.2}^{+4.8}$ & (0.05, 22.95) & log-uniform\\
Semimajor axis $a$ (AU)    &     ${15.7}_{-1.0}^{+3.5}$ & (14.1, 25.0) & $1/a$  (log-uniform) \\
$\sqrt{e} \sin \omega$\tablenotemark{\rm *}    &     ${-0.17}_{-0.44}^{+0.39}$ & (-0.66, 0.39) & uniform \\
$\sqrt{e} \cos \omega$\tablenotemark{\rm *}  &       ${0.22}_{-0.41}^{+0.24}$ & (-0.36, 0.55) & uniform  \\
Inclination ($^\circ$)    &     ${151.3}_{-12}^{+8.4}$   & (131, 165)& $\sin i$ (geometric) \\
PA of the ascending node $\Omega$ ($^\circ$)  &     ${281}_{-271}^{+69}$ & (1, 359) & uniform \\
Mean longitude at 2010.0 ($^\circ$) &        ${123}_{-71}^{+38}$ & (29, 190) & uniform \\
Parallax (mas)   &      ${24.544}_{-0.068}^{+0.068}$ & (24.41, 24.68) & Gaussian, $24.54 \pm 0.07$ \\
\hline
\multicolumn{4}{c}{Derived Parameters} \\ \hline
Period (yrs)     &    ${47.0}_{-4.5}^{+17}$ & (39.057, 95.567) \\
Argument of periastron $\omega$ ($^\circ$)\tablenotemark{\rm *}   &      ${251}_{-220}^{+75}$ & (4.659, 355.383) \\
Eccentricity $e$    &    ${0.28}_{-0.16}^{+0.13}$ & (0.023, 0.468) \\
Semimajor axis (mas)  &       ${386}_{-24}^{+85}$ & (345.307, 612.958) \\
Periastron time $T_0$ (JD)    &     ${2465218}_{-771}^{+3763}$ & (2459950.418, 2481296.372) \\
Mass ratio   &      ${0.0071}_{-0.0028}^{+0.0026}$ & (0.0, 0.013)
\enddata
\label{tab:mcmc_results-lognorm}
\end{deluxetable*}

  \begin{figure*}
    \begin{flushright}
    \includegraphics[width=1\textwidth]{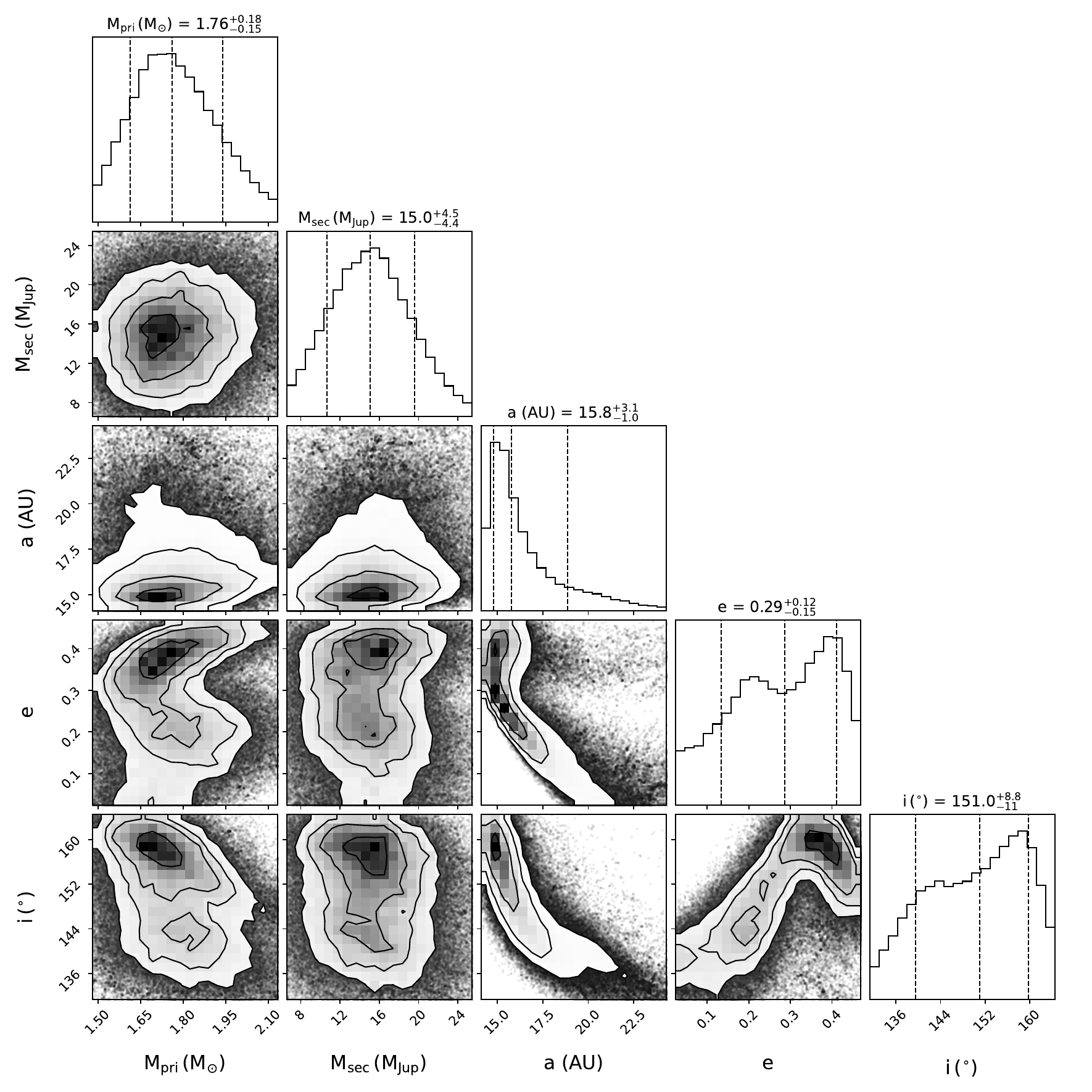} \\
    \vspace{-0.02\textwidth}
    \end{flushright}
    \caption{Corner plot showing MCMC posterior distributions for the flat prior. Fit using absolute astrometry from HGCA, relative astrometry (from both initial discovery and follow-up), and RV measurements from KPIC courtesy of \citet{Zhang2024}. The contours show the $68\%$ $(1\sigma)$, $95\%$ $(2\sigma)$, and $99\%$ $(3\sigma)$ confidence intervals.}
    \vspace{-0.in}
    \label{fig:corner}
\end{figure*} 

\begin{figure*}
   \includegraphics[width=0.34\textwidth,trim=0mm 3mm 2mm 0mm,clip]{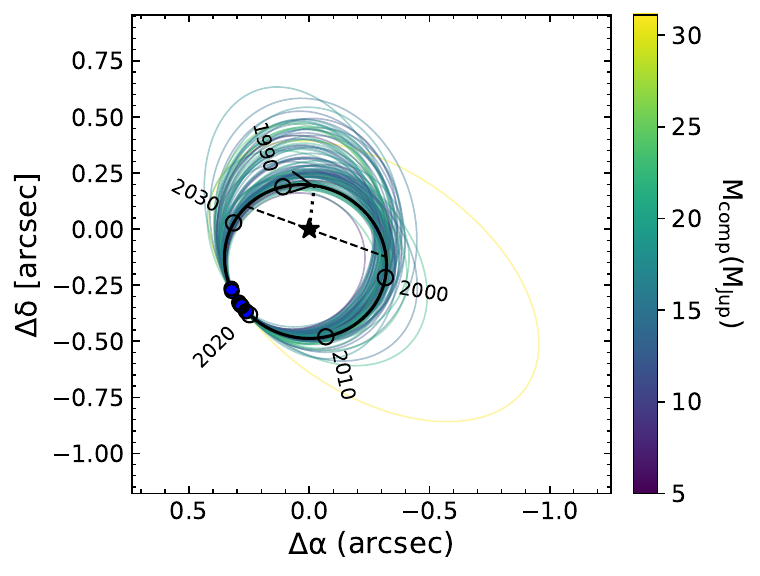}
    \includegraphics[width=0.32\textwidth,trim=0mm 3mm 2mm 0mm,clip]{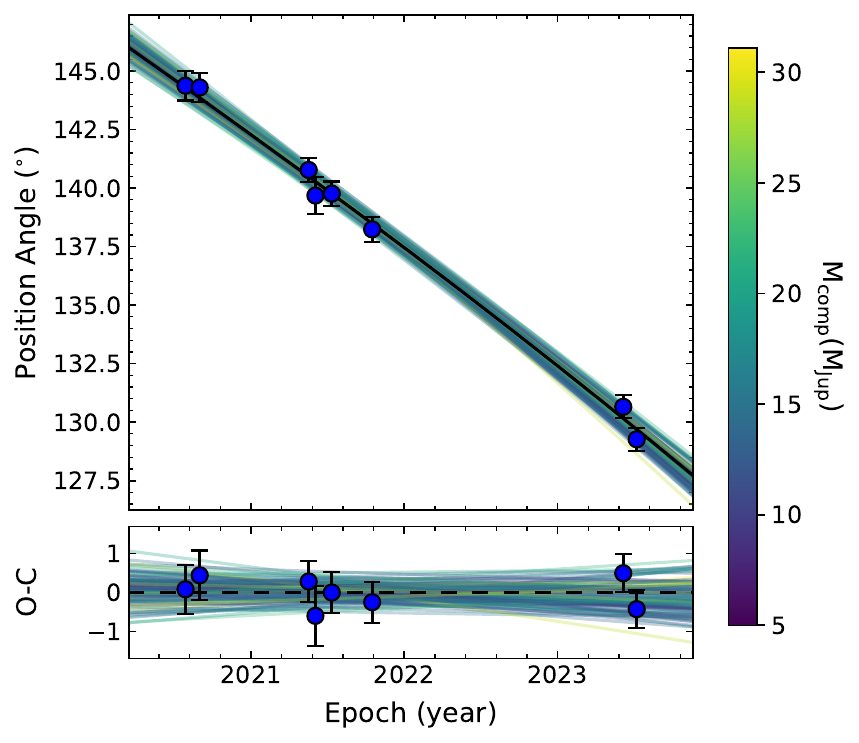}
    \includegraphics[width=0.32\textwidth,trim=0mm 3mm 2mm 0mm,clip]{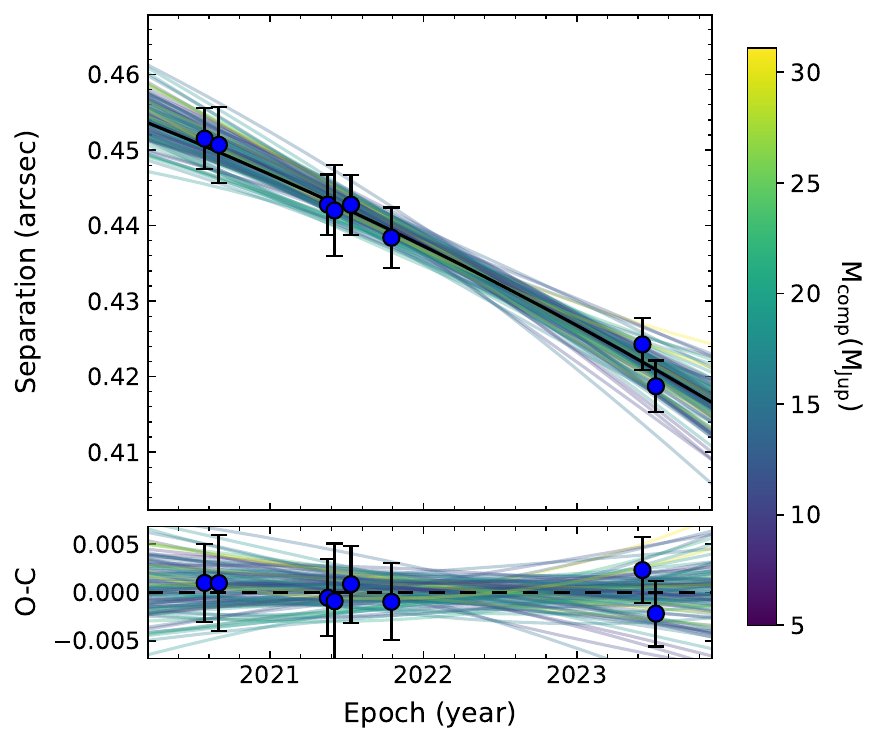}\\
     \vspace{-0.15in}
    \caption{Orbit fitting with \texttt{orvara} assuming flat (uniform) companion mass priors.  (Left) Predicted orbits with most likely orbit in black along with 100 randomly selected orbits (color coded by mass) from the MCMC posterior distribution. Blue circles show our data, while empty circles show predicted location per epoch. (Center) Position angle vs Epoch and (Right) Separation vs Epoch for the selected orbits.}
    \label{fig: orbit}
\end{figure*}

  \begin{figure*}
    \begin{flushright}
    \includegraphics[width=1\textwidth]{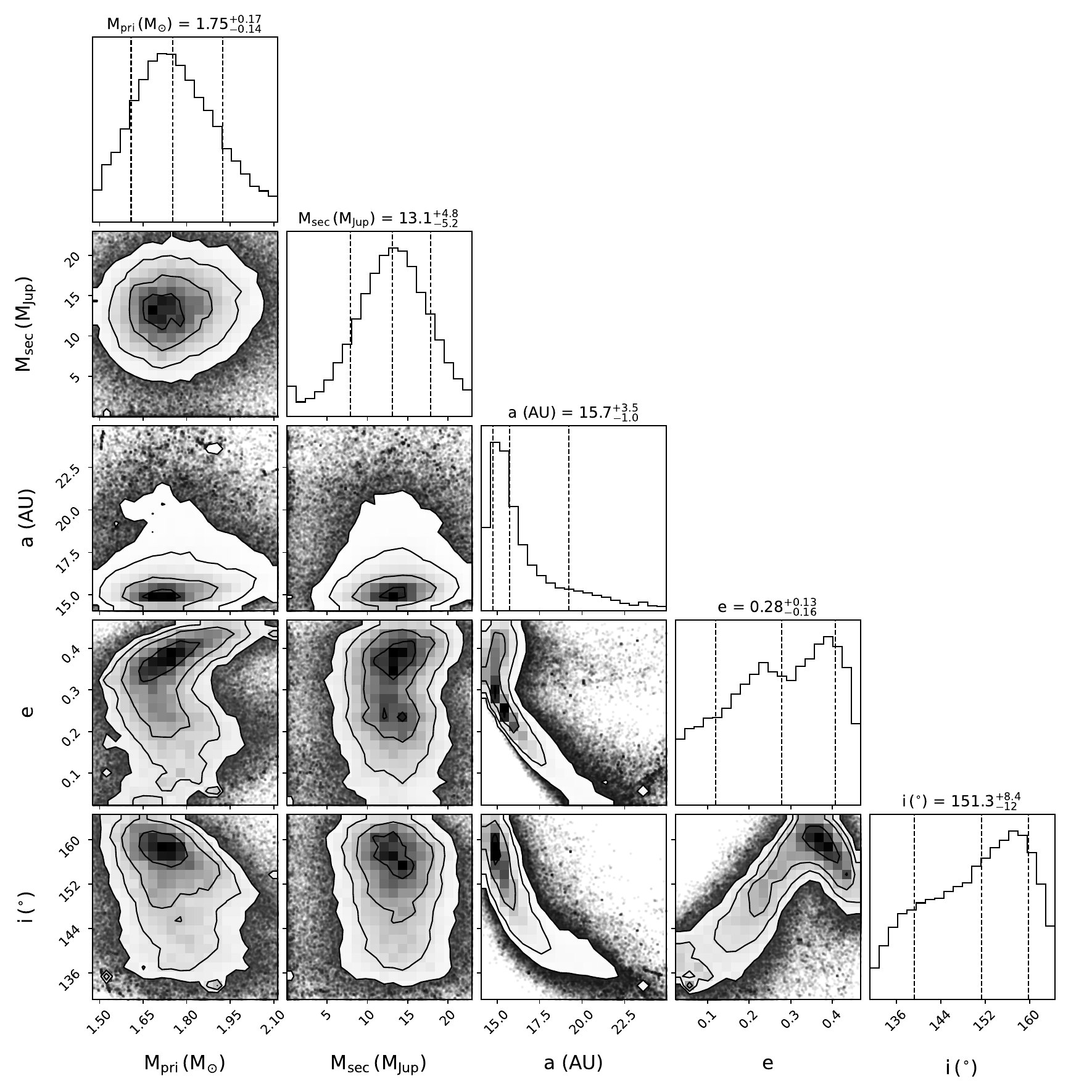} \\
    \vspace{-0.02\textwidth}
    \end{flushright}
    \caption{Corner plot showing MCMC posterior distributions for the log-uniform prior. Fit using absolute astrometry from HGCA, relative astrometry (from both initial discovery and follow-up), and RV measurements from KPIC courtesy of \citet{Zhang2024}. The contours show the $68\%$ $(1\sigma)$, $95\%$ $(2\sigma)$, and $99\%$ $(3\sigma)$ confidence intervols.}
    \vspace{-0.in}
    \label{fig:corner_log-norm}
\end{figure*} 

  \begin{figure*}
    \begin{flushright}
    \includegraphics[width=1\textwidth]{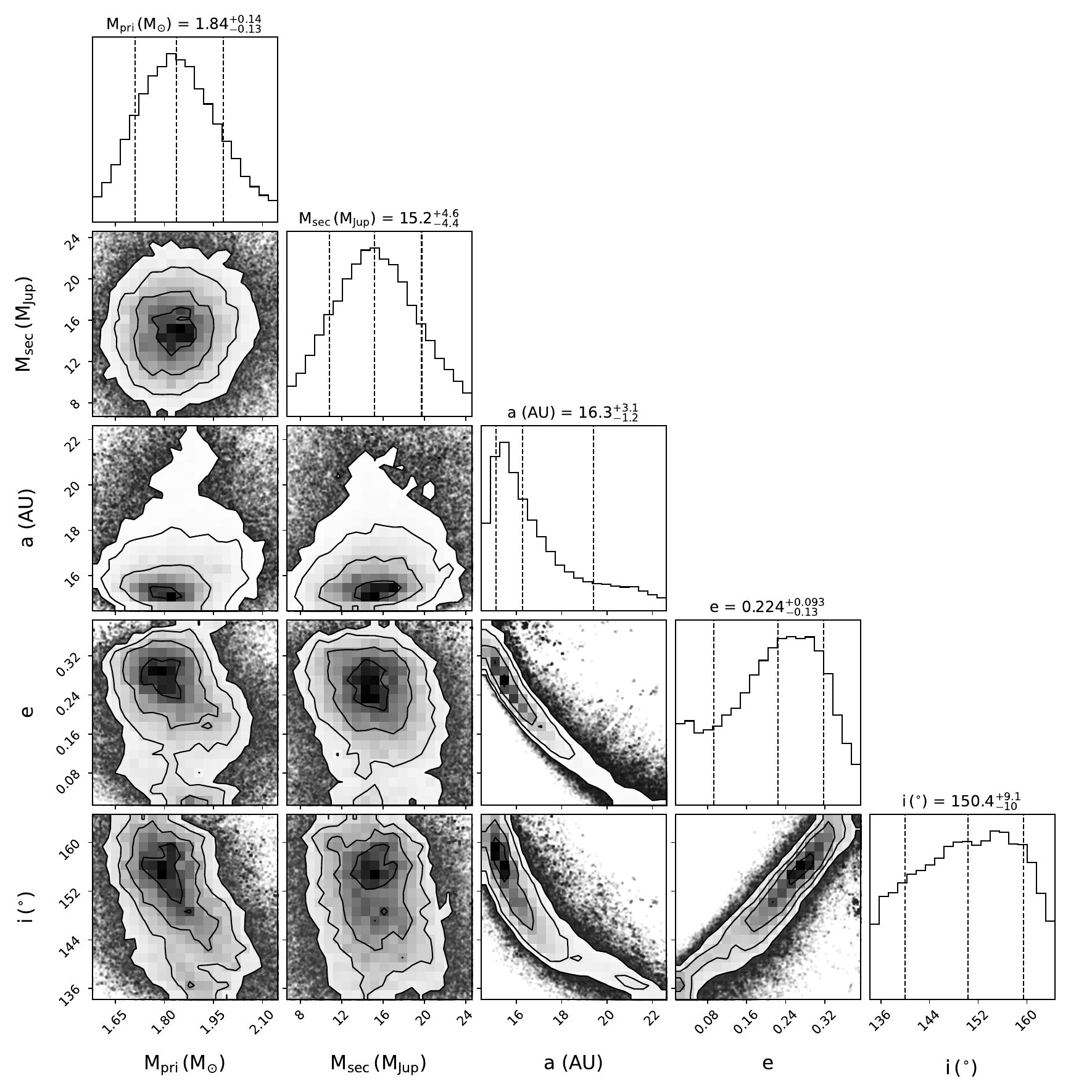} \\
    \vspace{-0.02\textwidth}
    \end{flushright}
    \caption{Same as Figure \ref{fig:corner} except including the May 2023 VLTI/GRAVITY astrometric point from \citet{Winterhalder2025}.}
    \vspace{-0.in}
    \label{fig:corner_flatvlt}
\end{figure*} 

  \begin{figure*}
    \begin{flushright}
    \includegraphics[width=1\textwidth]{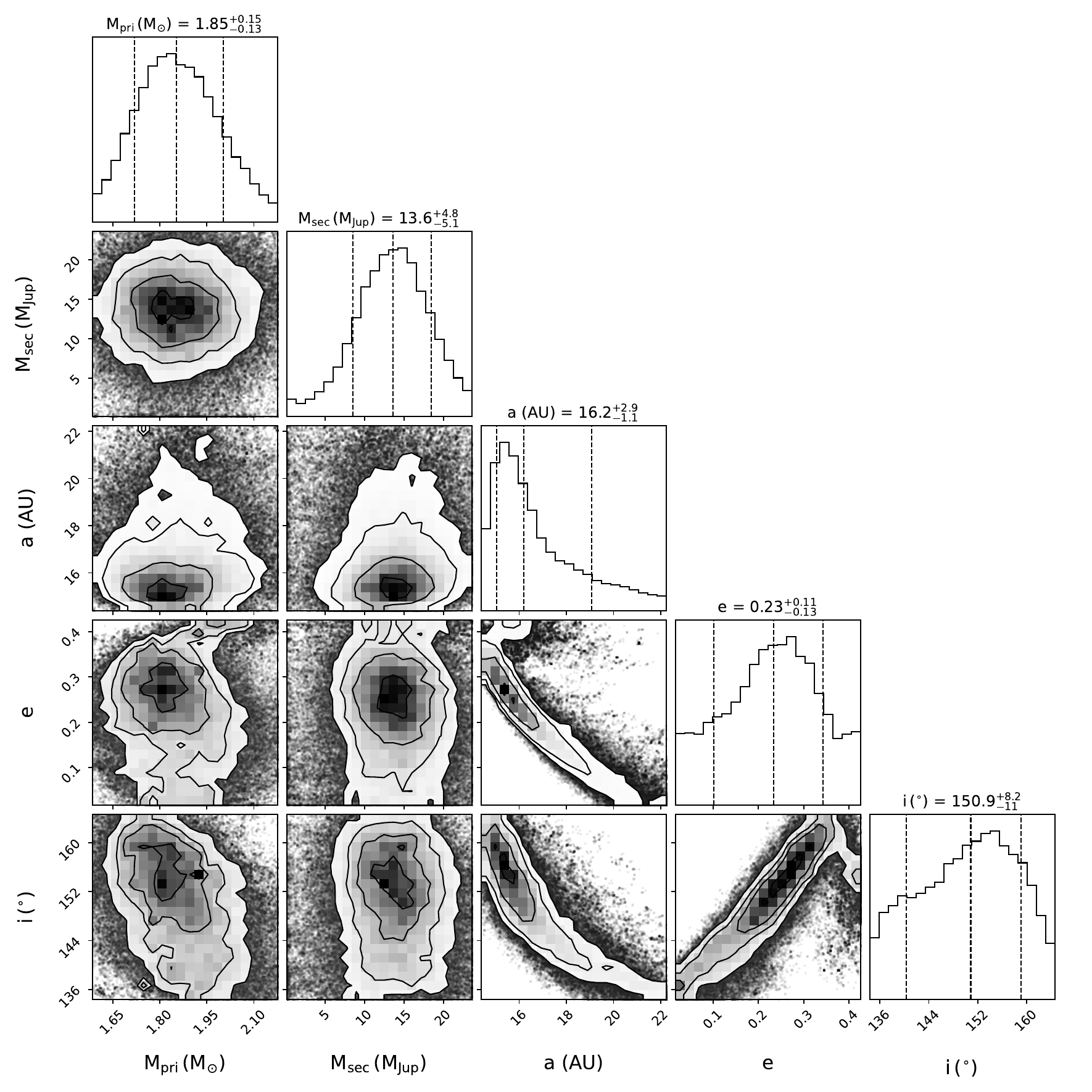} \\
    \vspace{-0.02\textwidth}
    \end{flushright}
    \caption{Same as Figure \ref{fig:corner_log-norm} except including the May 2023 VLTI/GRAVITY astrometric point from \citet{Winterhalder2025}.}
    \vspace{-0.in}
    \label{fig:corner_log-normvlt}
\end{figure*}

\begin{figure}
\centering
   \includegraphics[width=0.45\textwidth,trim=0mm 6mm 5mm 4mm,clip]{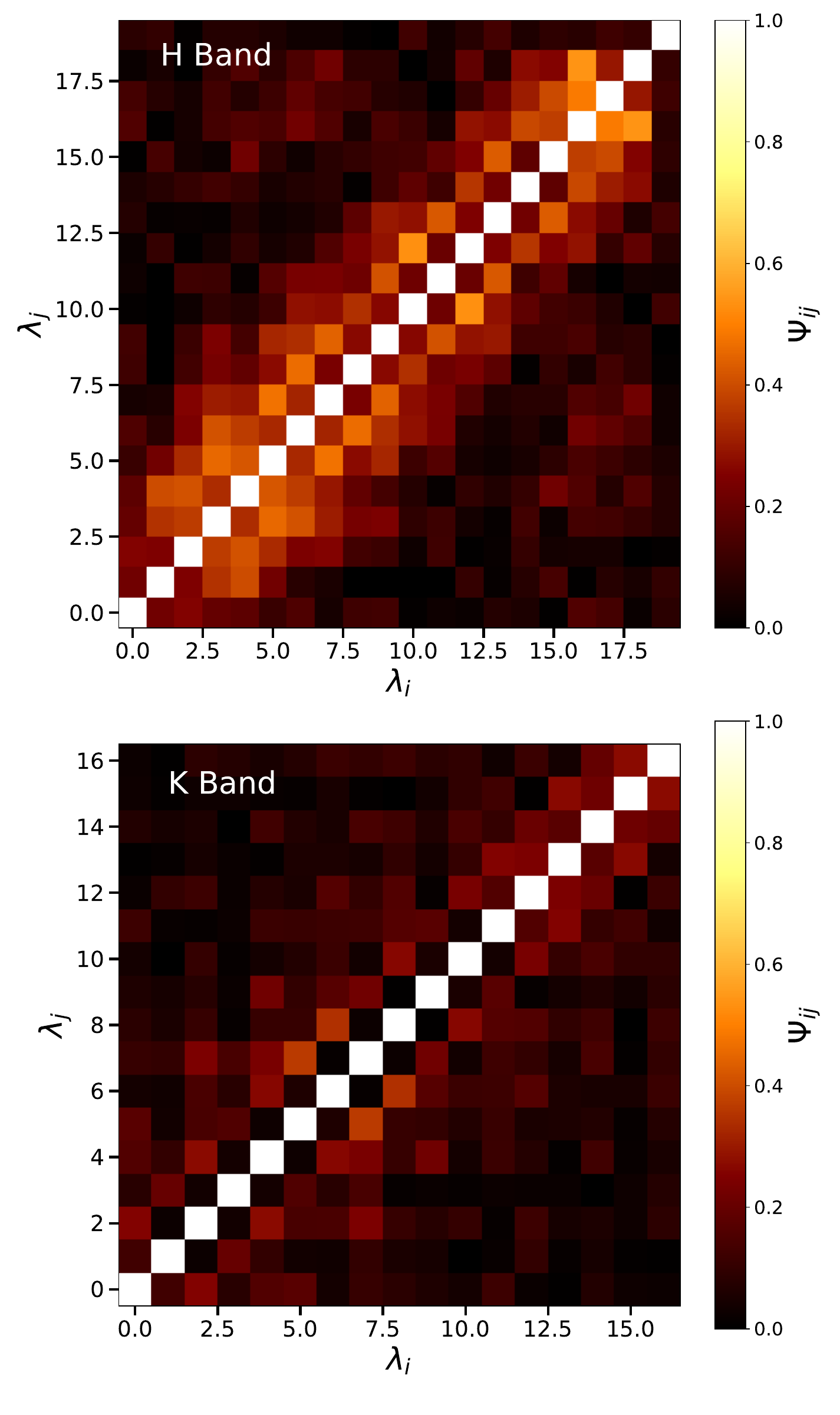}
   \caption{Correlation Matrix $\Psi_{i,j}$ for H (top) and K Band (bottom) as a function of spectral chanel $\lambda_i, \lambda_j$. The off-diagonal elements are residual spectrally correlated noise.}
   \label{fig:covar} 
\end{figure}

\begin{figure*}
    \includegraphics[width=0.48\textwidth,trim=4mm 4mm 2mm 0mm,clip]{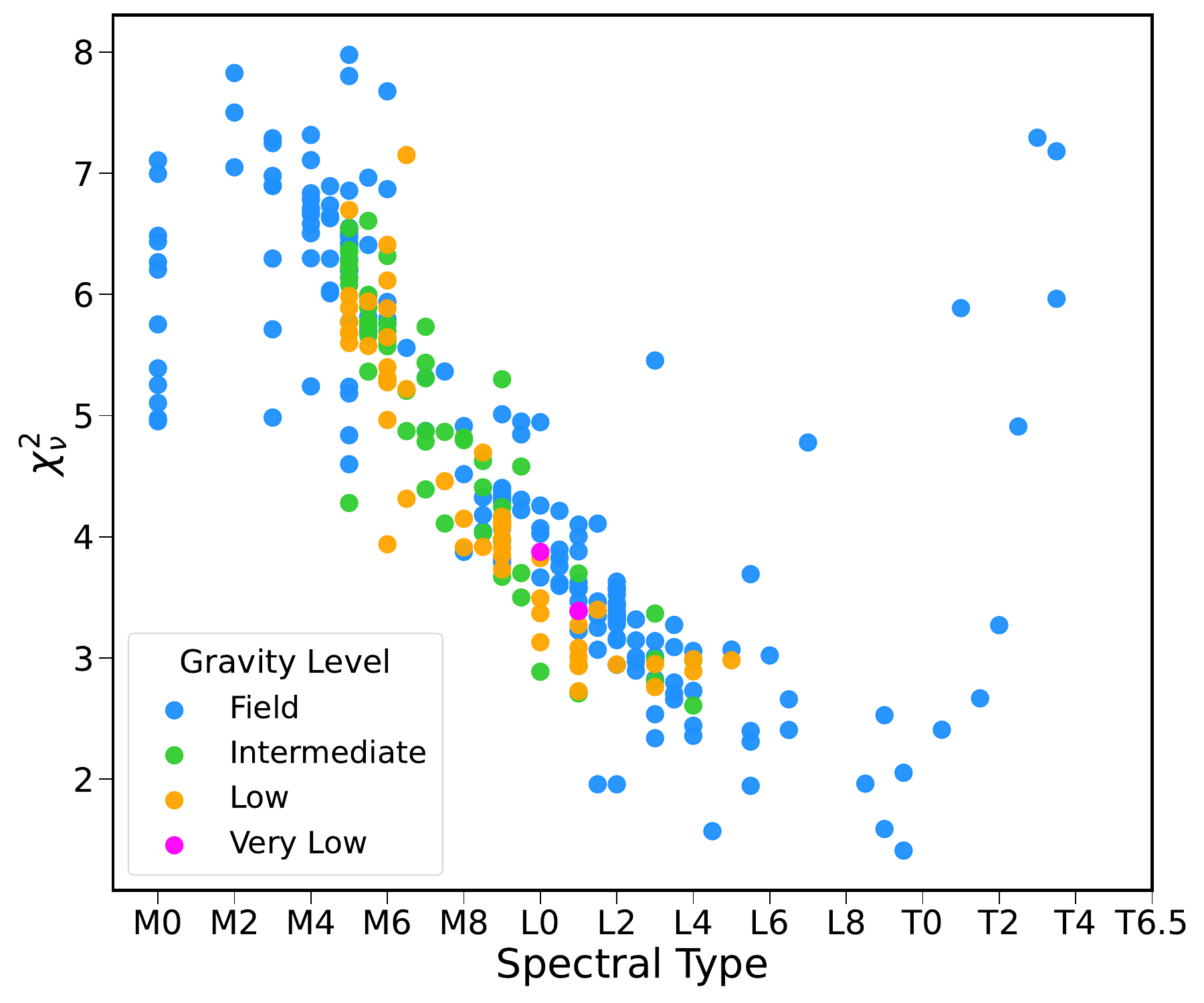}
    \includegraphics[width=0.48\textwidth,trim=4mm 3mm 2mm 0mm,clip]{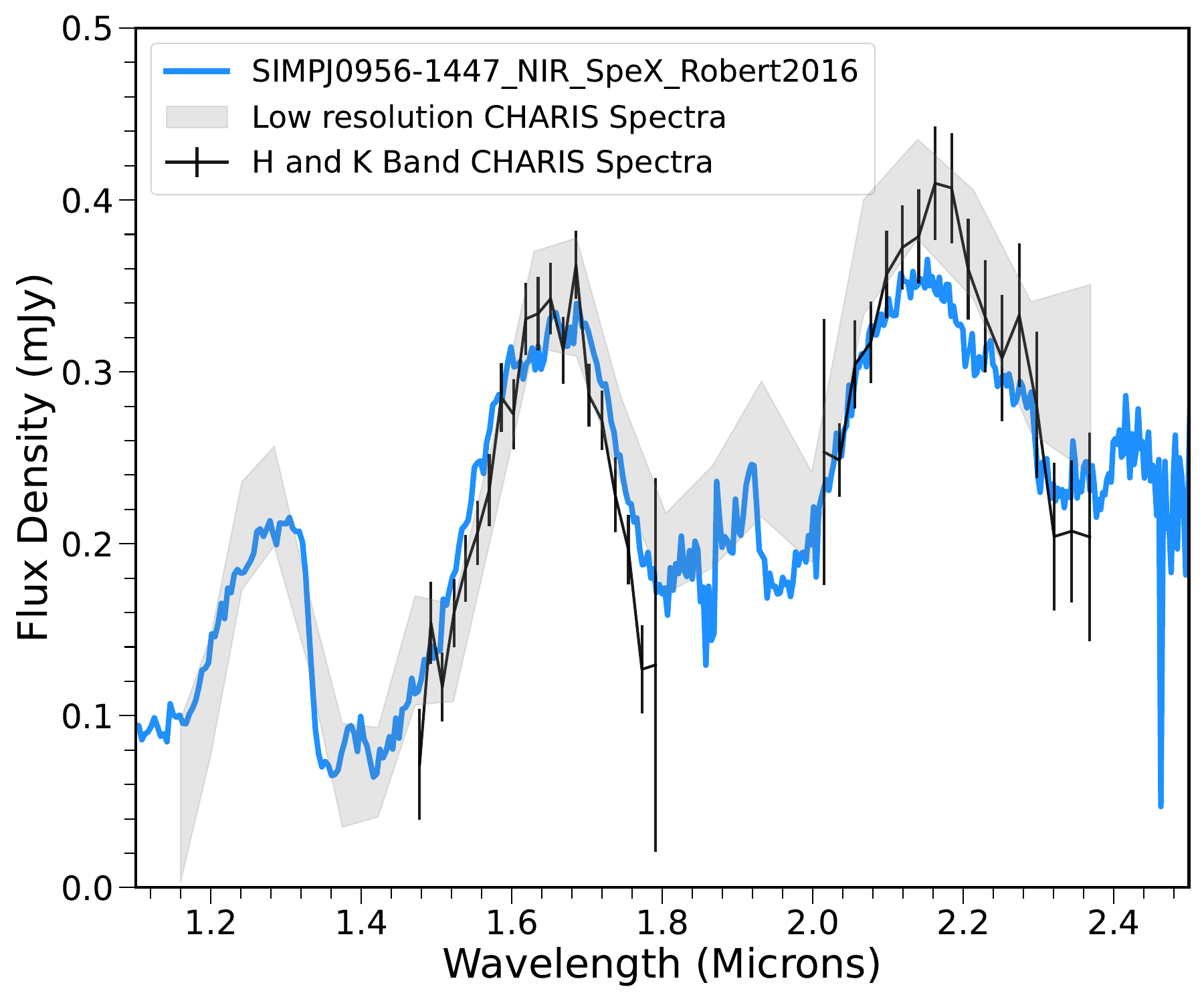}\\
     \vspace{-0.15in}
    \caption{(Left) $\chi_{\nu}^2$ versus spectral type for objects in the Montreal Spectral Library \citep{Gagne2014} when compared to HIP 99770 b H and K band spectra. The sources are colored based on their gravity level. The best fitting objects fall in the L4.5 to L9.5 spectral type range. This spectral range is consistent with a temperature of $1300K-1600K$. (Right) The best fitting spectra from the Montreal Spectral Library, SIMPJ0956-1447, shown in blue compared to the HIP 99770 b H and K band spectra shown in black, with the low resolution spectra plotted in gray as a reference for shorter wavelengths.}
    \label{fig: MSL}
\end{figure*}

\section{Astrometric Analysis}
Using the \texttt{orvara} Markov Chain Monte Carlo-based dynamical code \citep{Brandt2021b}, we jointly model the host star's absolute astrometry from HGCA and the companion’s relative astrometry from both the discovery paper and our H and K band follow-up (Table \ref{tab:astrom}) to constrain the companion’s orbital parameters. 
To these data, we add the companion relative RV estimate of 2.6 $\pm$ 0.5 km s$^{-1}$ from KPIC data \citep{Zhang2024}. The new relative astrometric measurements more than double the available astrometric baseline for HIP 99770 b, while the relative RV measurement provides an additional constraint on HIP 99770 b's dynamical influence beyond the proper motion anomaly it induces on the host star.

We ran two separate simulations with \texttt{orvara}, one with a flat prior\footnote{Following \citet{Currie2023b}, we model a flat prior with \texttt{orvara} as a gaussian prior centered on 1 $M_{\rm Jup}$ with a standard deviation of 1000 $M_{\rm Jup}$.  This is equivalent to a true flat prior to within 1\% for masses less than 200 $M_{\rm Jup}$: well within the relevant companion mass range for HIP 99770 b.} and one with a log-uniform prior on the secondary mass with limits of 10$^{-4}$ to 10$^{3}$ $M_{\rm Jup}$. The different priors can be connected to assumptions about the underlying frequency of companions as a function of mass over the range relevant for HIP 99770 b. For planets $\lesssim$ 25$M_{\rm Jup}$, the mass function decreases with increasing masses, while at companion masses of $\approx$50 $M_{\rm Jup}$--80 $M_{\rm Jup}$ the companion frequency slightly increases with mass \citep[e.g.][]{Sahlmann2011,Kiefer2019,Currie2023b}. The mass at which the companion function turns over (i.e. transitions from decreasing with mass to increasing) varies in the literature. However, the reported range at which this turnover occurs covers $\approx$25--40 $M_{\rm Jup}$ \citep[][see also \citealt{Grether2006,Ma2014,Kiefer2019}]{Currie2023b}. As in \citet{Currie2023b}, we perform fits both with a flat prior on the companion mass and with a log-uniform prior, where the former provides a more conservative assumption, raising the possibility that we may be overestimating the mass.

For both simulations, we run \texttt{orvara} with 20 temperatures, with 100 walkers for each temperature, and 200,000 steps per walker. The first 1500 steps are burn-in and the chain is thinned by a factor of 50. The right columns of Table \ref{tab:mcmc_results} and \ref{tab:mcmc_results-lognorm} list the priors for each parameter for the simulations. 

\subsection{Results for A Flat Companion Mass Prior}
Figure \ref{fig:corner} shows the resulting posterior distributions for the flat prior on the companion mass: these values and other derived orbital properties are summarized in Table \ref{tab:mcmc_results} (middle columns), while Figure \ref{fig: orbit} displays predicted orbits for different solutions. We find primary and companion masses of $M_{\rm pri}$ = $1.76_{-0.15}^{+0.18}$ $M_{\odot}$ and $M_{\rm sec}$ = $15.0_{-4.4}^{+4.5}$ $M_{\rm Jup}$. The posterior mean and 68\% confidence intervals for the semimajor axis, eccentricity, and inclination are $a=15.8_{-1.0}^{+3.1}au$, $e=0.29_{-0.15}^{+0.12}$, and $i=151.0_{-11}^{+8.8}$ \textdegree, respectively. The mass ratio $q = 0.008_{-0.002}^{+0.003}$.

Most parameters plotted have single-peaked posterior distributions, although the eccentricity posterior mean appears to straddle two peaks of $e$ $\sim$ 0.225 and 0.375. The eccentricity is shifted slightly higher than what was found in \citet{Currie2023b}, disfavoring a circular orbit. The corner plot shows slight covariances between the eccentricity and inclination vs. semimajor axis, where high eccentricities and more face-on orbits appear to favor smaller orbits. Unlike \citet{Currie2023b}, the companion mass appears uncorrelated with the inclination.  

Generally speaking, the posterior means agree with those from \citet{Currie2023b}, while the 68\% confidence intervals are smaller.   
Compared to the discovery paper mass estimates, the companion mass posterior peak is $\sim$7\% smaller with $\pm$1-$\sigma$ range that is $\sim$15\% smaller, although it is formally consistent with prior results. Because the primary mass posterior distribution is also shifted towards smaller values by 5\%, the mean of the companion-to-star mass ratio is essentially identical to that from \citet{Currie2023b}. The semimajor axis posterior distribution also noticeably shifts to smaller values by the same amount as the companion mass but is formally consistent with previous results. The uncertainties for the semimajor axis, eccentricity, and inclination are 21\%, 17\%, and 10\% smaller.

\subsection{Results for A Log-Uniform Companion Mass Prior}

Figure \ref{fig:corner_log-norm} shows the resulting posterior distributions for the log-uniform prior on the companion mass; Table \ref{tab:mcmc_results-lognorm} (middle columns) summarizes the results. The mean of the companion mass posterior distribution drops to $M_{sec}=13.1_{-5.2}^{+4.8} M_{\rm Jup}$, about 6\% lower than the value found in \citet{Currie2023b}. The primary mass posterior distribution is nearly identical to that for the flat companion mass prior, leading to a small mass ratio of $q = {0.0071}_{-0.0028}^{+0.0026}$. All other orbital parameter posteriors are nearly identical to their values from the flat prior.

\subsection{Results Including Recent VLTI/GRAVITY Astrometry}

During the final preparation of this work, \citet{Winterhalder2025} published VLTI/GRAVITY astrometry of HIP 99770 b at a similar epoch to our CHARIS $H$ and $K$ band data.  The high precision of GRAVITY astrometry ($\sim$ 0.1 mas) have often resulted in tighter constraints on planet orbits \citep[e.g.][]{Balmer2025}.
To assess the impact of the GRAVITY data on our results, we re-ran both \texttt{orvara} simulations including the May 2023 astrometry from \citet{Winterhalder2025}.   Figures \ref{fig:corner_flatvlt} and \ref{fig:corner_log-normvlt} show resulting posterior distributions assuming a flat and log-uniform prior, respectively. 

Including the VLTI/GRAVITY data causes the secondary eccentricity peak at $e$ $\sim$ 0.4 to disappear. 
  The eccentricity distribution is more narrowly constrained, albeit at smaller median posterior values compared to those found by \citet{Winterhalder2025} using a different dynamical code and without incorporating KPIC relative RV data:  ($e$ = ${0.224}_{-0.13}^{+0.093}$ and ${0.23}_{-0.13}^{+0.11}$ for the flat prior and log-uniform prior, respectively).   Otherwise, including GRAVITY data has negligible effects on our parameter confidence intervals: HIP 99770 b's dynamical mass is the same as found from simulations without including GRAVITY data to within 1--4\% and the semimajor axes and inclinations are likewise well within agreement.

\subsection{The Role of HIP 99770 b's Relative RV Measurement from KPIC}
While additional relative astrometry may improve estimates of most orbital parameters, the systematically lower companion mass estimates are almost entirely due to the relative RV measurement from \citet{Zhang2024}. To illustrate this point, we ran \texttt{orvara} with our additional relative astrometry but without the KPIC RV measurement for a flat companion mass prior (Figure \ref{fig:corner_norv}). Like other simulations described above, our results find a better constrained semimajor axis, eccentricity, and inclination. However, the companion mass posterior is higher, $16.1_{-4.7}^{+5.7}$ $M_{\rm Jup}$, 
and nearly indistinguishable from the same calculation in \citet{Currie2023b}. The primary mass is higher and more similar to previously published results.

\section{Spectroscopic Analysis}

\subsection{Spectral Covariance}

Figure \ref{fig:covar} shows the spectral covariance matrices \citep{GrecoBrandt2016} for the H and K passbands. The spectral covariance matrices take into account spatially and spectrally correlated noise. The off-diagonal elements are spectrally correlated noise. To find the average spectral correlation we use the functional form from \citet{GrecoBrandt2016}:

\begin{multline}
    \psi_{ij}\approx A_{\rho} \exp \left[-\frac{1}{2}\left(\frac{\rho}{\sigma_{\rho}}\frac{\lambda_i-\lambda_j}{\lambda_c}\right)^2\right] \\
    +A_{\lambda} \exp \left[-\frac{1}{2}\left(\frac{1}{\sigma_{\lambda}}\frac{\lambda_i-\lambda_j}{\lambda_c}\right)^2\right] + A_{\delta}\delta_{ij}.
    \label{eq:covar}
\end{multline}

Here, $\rho$ is the angular distance from the star, $\lambda$ is the wavelength, A is the amplitude of the noise, $\sigma$ is the correlation length that characterize the noise, and $\delta_{ij}$ is the Kronecker delta. 

For H we find $A_{\rho}\sim 0.30$, $A_{\lambda}\sim 0.07$, $A_{\delta}\sim 0.61$, $\sigma_{\rho}\sim 0.32$, and $\sigma_{\lambda}\sim 2.90$. Thus, about 60\% of the noise at HIP 99770 b's angular separation is uncorrelated with wavelength and angular separation: the noise for neighboring channels reaches $\psi$ $\sim$ 0.5. For K band, the noise is less correlated ($A_{\delta}\sim 0.81$, $A_{\rho}\sim 0.06$, $A_{\lambda}\sim 0.11$, $\sigma_{\rho}\sim 0.89$, and $\sigma_{\lambda}\sim 0.03$).

\subsection{Empirical Constraints on HIP 99770 b's Atmosphere}
\subsubsection{Spectral Type}

To empirically estimate HIP 99770 b's spectral type
we compare its H and K band spectra to over 300 substellar objects in the Montreal Spectral Library \citep[MTL][]{Gagne2014}. The MTL library includes M, L, and some T dwarfs with a range of gravity classifications. We bin down the MTL spectra to CHARIS's resolution and interpolate flux densities onto the CHARIS wavelength arrays. To estimate the agreement between the $jth$ library spectrum and HIP 99770 b's data, we compute the $\chi^{2}$ statistic as:
\begin{equation}
    \chi_{j}^{2} = R_{H, j}^{T}C_{H}^{-1}R_{H,j} + R_{K, j}^{T}C_{K}^{-1}R_{K,j},
    \label{eq:chi2emp}
\end{equation}
where $R_{H,j}$ and $R_{K,j}$ equal $f_{spec} -\alpha_{j} F_{spec}$: the difference between the HIP 99770 b's flux density and the library spectrum, scaled by a factor of $\alpha$ to minimize $\chi^{2}$. $C_{H}$ and $C_{K}$ are the covariances for each passband.

The best-matching spectrum is the field L9.5 dwarf SIMPJ0956-1447 ($\chi^{2}_{\nu}$ $\sim$ 1.4; Figure \ref{fig: MSL}, Left). This object also best-reproduced CHARIS spectra for HD 33632 Ab, albeit with a smaller $\chi^{2}$ value \citep[$\chi^{2}_{\nu}$ $\sim$ 1.1;][]{ElMorsy2024}. While the L4.5 dwarf SIMPJ1122+0343 is formally the next best fitting object ($\chi^{2}_{\nu}$ $\sim$ 1.6), its low $\chi^{2}$ value is likely an artifact of its noisier spectrum (SNR $\approx$ 10 per channel).
Considering the entire set of MTL objects, we find a minimum near $\sim$L6-L9.5, with a slight preference for L8--L9.5 objects. From \citet{Stephens2009}, the L6-L9.5 spectral type range favors temperatures of $\approx$1300-1500 $K$.

\subsubsection{Comparisons of HIP 99770 b's Spectral Shape with Other L/T Transition Objects}

The MTL library sparsely samples the range of gravity classifications near the L/T transition, which may also correlate with various levels of disequilibrium chemistry \citep[e.g.][]{Barman2011a}. To explore gravity and chemistry for HIP 99770 b, we first compare its CHARIS $H$ and $K$-band spectra to the best-fit field object SIMPJ0956-1447 and directly imaged substellar companions with direct dynamical mass measurements or limits: HR 8799 bcd and HD 33632 Ab \citep{Marois2008a,Currie2020a}. HR 8799 b and HR 8799 cd have best-estimated masses of $\approx$5--6 $M_{\rm Jup}$ and 7--10 $M_{\rm Jup}$ \citep{Wang2018}\footnote{Formally, only HR 8799 e has a direct dynamical mass measurement from imaging and astrometry \citep{GBrandt2021b}. HR 8799 bcd's mass constraints in \citet{Wang2018} come from dynamical stability arguments (yielding a mass upper limit) coupled with luminosity evolution models (giving a mass lower limit).}. HD 33632 Ab's mass is ${52.8}_{-2.4}^{+2.6}$$M_{\rm Jup}$ \citep{ElMorsy2024}. SIMPJ0956-1447 and HD 33632 Ab are field objects with likely ages in excess of 1 Gyr, while HR 8799 is likely $\approx$40 Myr old \citep{Baines2012,Robert2016,GBrandt2021}.

For HD 33632 Ab, we adopt $H$ and $K$ band spectra from \citet{ElMorsy2024}.  There are various sources for HR 8799 bcd spectra. In H band, we adopt HR 8799 b spectra from Keck/OSIRIS \citep{Barman2011a} and GPI spectra for HR 8799 c from \citet{Greenbaum2018}. In K band, we use VLTI/GRAVITY spectra from \citet{Nasedkin2024} for HR 8799 cd, using the OSIRIS spectrum for the b planet as a comparison.

\begin{figure*}
    \includegraphics[width=0.48\textwidth,trim=15mm 3mm 30mm 20mm,clip]{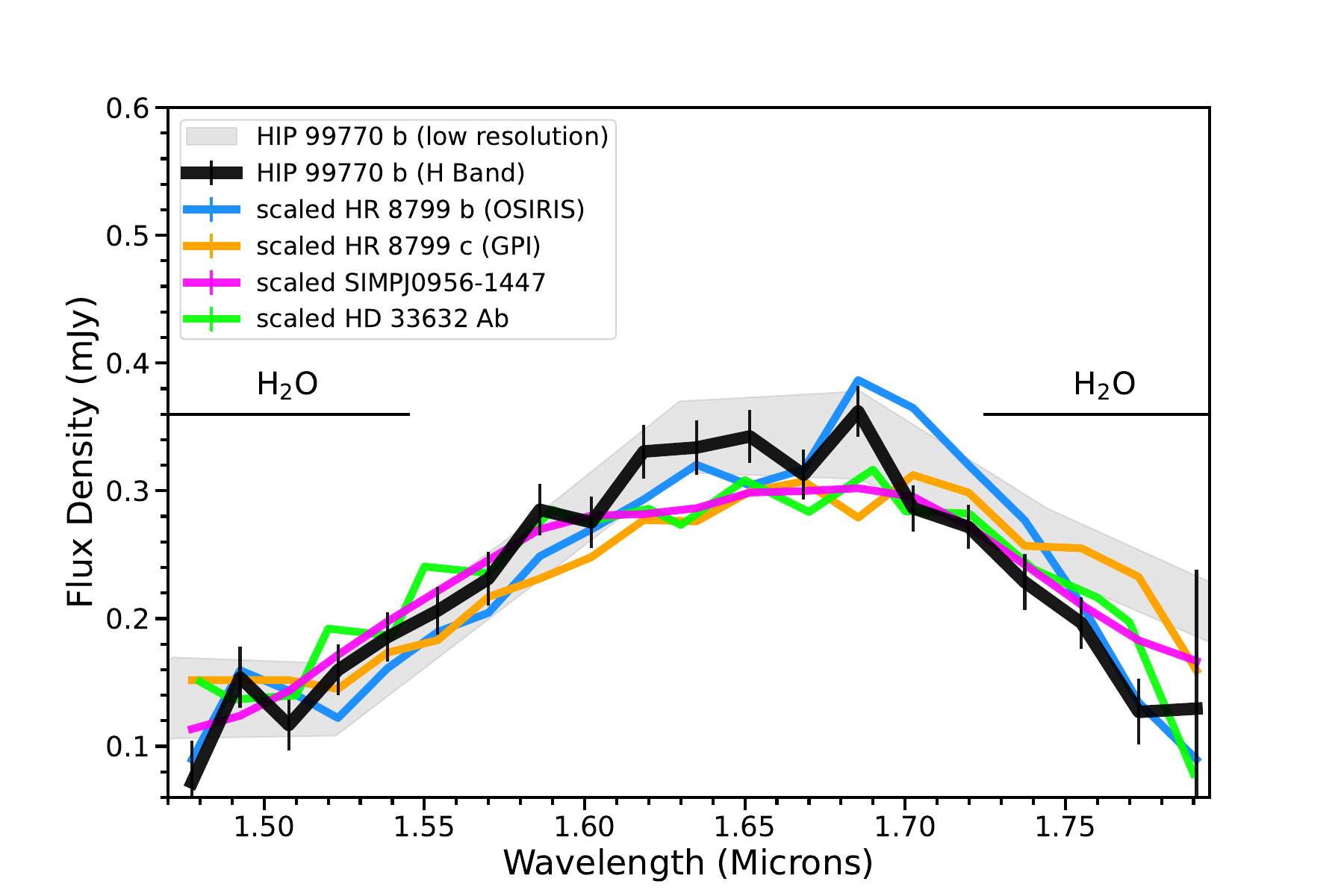}
    \includegraphics[width=0.48\textwidth,trim=10mm 3mm 30mm 22mm,clip]{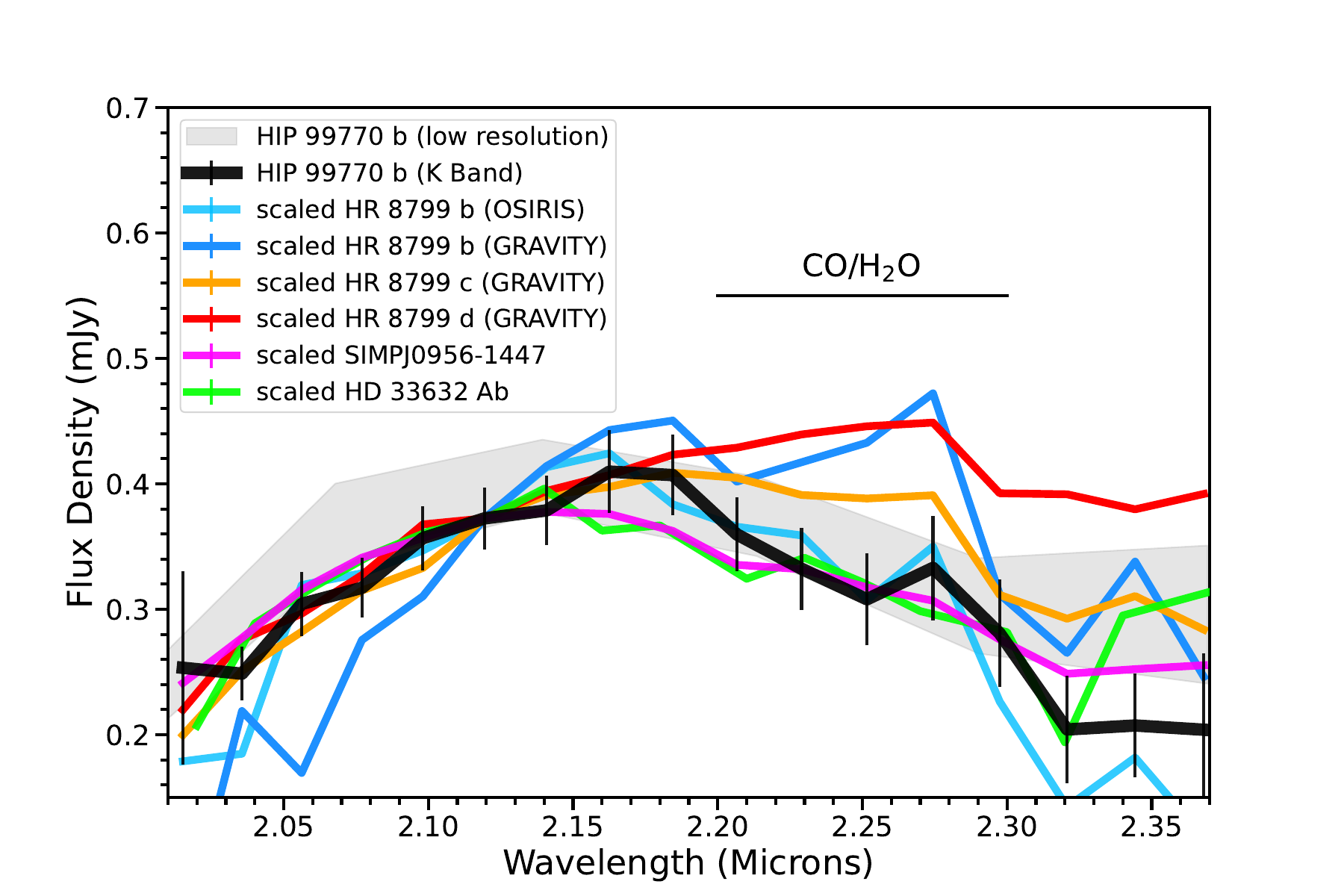}\\
     \vspace{-0.15in}
    \caption{(Left) HIP 99770 b H band spectra (black) compared to the H band spectra of HR 8799 b (blue), HR 8799 c (orange), SIMPJ0956-1447 (magenta), and HD 33632 Ab (green). The low resolution spectra is plotted in gray and H$_2$O absorption is also marked in black. (Right) HIP 99770 b K band spectra (black) compared to the K band spectra of HR 8799 b (light blue: OSIRIS, dark blue: GRAVITY), HR 8799 c (orange), HR 8799 d (red), SIMPJ0956-1447 (magenta), and HD 33632 Ab (green). For both plots the low resolution spectra are plotted in gray and H$_2$O and CO absorption is also marked in black.}
    \label{fig: Comp}
\end{figure*}

Figure \ref{fig: Comp} compares HIP 99770 b’s H (Left) and K (Right) band spectrum to spectra from SIMPJ0956-1447, HD 33632 Ab, and HR 8799 bcd normalized to match HIP 99770 b's mean flux density (H band) or its measurement at 2.16 $\mu m$ (K band).
In H band, older field objects SIMPJ0956-1447 and HD 33632 Ab match HIP 99770 b's spectrum well at the shortest and longest wavelengths affected by water opacity but have a more flattened peak at 1.6--1.7 $\mu m$. HR 8799 bc are redder -- less emission than HIP 99770 b at 1.5--1.6 $\mu m$ and more at $\lambda$ $\gtrsim$ 1.7 $\mu m$ -- although the HR 8799 c has a flattened peak emission.  

At K band, HIP 99770 b's triangular-shaped spectrum is well matched by HD 33632 Ab and SIMPJ0956-1447. At the reddest K-band wavelengths sensitive to carbon monoxide absorption and thus disequilibrium carbon chemistry, the spectra of HR 8799 c and especially d are far flatter with reduced CO absorption compared to HIP 99770 b and the field objects.   
The interpretation of HIP 99770 b vs. HR 8799 b is less clear.   
Although their reported spectroscopic errors at CHARIS's resolution are small (e.g. typically $\approx$10\% for OSIRIS), the VLTI/GRAVITY spectrum and Keck/OSIRIS spectrum spectra have very different shapes.
The GRAVITY HR 8799 b spectrum is flatter, also consistent with greater disequilibrium chemistry than HIP 99770 b and previous result at longer wavelengths \citep[e.g.][]{Skemer2012}, while the OSIRIS spectrum is more similar to HIP 99770 b's.

\subsubsection{Surface Gravity}
\begin{deluxetable*}{lllllllll}
     \tablewidth{0pt}
     \tabletypesize{\scriptsize}
    \tablecaption{Gravity Classification Based on Spectral Indices\label{tab:indices}}
    \tablehead{\colhead{Object} & \colhead{H-cont} &  \colhead{HPI} &\colhead{TLI-g} & \colhead{WH} & \colhead{CH$_{\rm 4}$-1.6} & \colhead{CH$_{\rm 4}$-K} & \colhead{H$_{2}$(K)} & {Average Score/Classification}}
    \startdata
   HIP 99770 b & 0.76 (0) &  2.33 (3)& 0.75 (3)& 0.75 (2--3)& 0.83 (2--3) & 0.85 (0) & 1.36 (0) & 1.57/$\beta$--$\gamma$\\
    HR 8799 b & 0.90 (1) &  1.98 (3) & 0.72 (3)& 0.72 (2--3)& 0.79 (3) & 1.16 (3) & 1.12 (0) & 2.21/$\gamma$\\
    HR 8799 c & 1.13 (2) & 1.56 (0--1)&  0.84 (2--3) & 0.74 (2--3)& 0.82 (2--3) & 1.03 (1) & 1.10  (1?) & 1.71/$\beta$--$\gamma$\\
    SIMPJ0956-1447& 0.81 (0) & 1.70 (1) & 0.93 (0--1)& 0.88  (0--1)& 0.94 (0--1)& 0.82 (0) & 1.22 (0) & 0.36/field\\
    HD 33632 Ab & 0.83 (0) & 1.73 (2) & 0.90 (0--1)& 0.91 (0--1)& 0.98 (0) & 0.81 (0) & 1.17 (0) & 0.43/field\\
    \enddata
    \tablecomments{In parentheses, we list the gravity score for each object, following \citet{AllersLiu2013} as modified in this paper: 0 for field, 1 for intermediate gravity ($\beta$), 2 for low gravity ($\gamma$), and 3 for very low gravity ($\delta$).}
    \end{deluxetable*}

Because there are very few MTL mid L to early T dwarfs with intermediate ($\beta$) or young ($\gamma$) gravity classifications, we use spectral indices from \citet{Piscarreta2024} and \citet{AllersLiu2013} to assess HIP 99770 b's gravity classification, comparing the results to HR 8799 bcd, HD 33632 Ab, and SIMPJ0956-1447.  We compute the H band continuum index from \citet{AllersLiu2013}, the HPI (``H peak"), TLI-g, WH, CH$_{\rm 4}$-1.6, and CH$_{\rm 4}$-K indices from \citet{Piscarreta2024}, and the H$_2$(K) index from \citet{Canty2013}.

 By convention, each index is computed in $F_{\rm \lambda}$ space.  The H continuum index is 
\begin{equation}
    \text{H-cont}=\left( \frac{\lambda_{l}-\lambda_{c1}}{\lambda_{c2}-\lambda_{c1}}F_{c2}+\frac{\lambda_{c2}-\lambda_{l}}{\lambda_{c2}-\lambda_{c1}}F_{c1}\right)/F_{l},
    \label{eq:hcont_index}
\end{equation}

\noindent where $\lambda_l$= 1.560 $\mu$m is the line wavelength, $\lambda_{c1}=1.470$ $\mu$m and $\lambda_{c2}=1.670$ $\mu$m are the continuum wavelengths, and the corresponding flux densities ($F_l$, $F_{c1}$, and $F_{c2}$) are in $F_\lambda$ units \citep{AllersLiu2013}. 

Other indices are flux ratios. E.g. the H$_{2}$(K) index from \citet{Canty2013} is
\begin{equation}
    \text{H$_2$(K)}=\frac{F_{\lambda}(2.17 \mu m)}{F_{\lambda}(2.24 \mu m)}, 
    \label{eq:h2k_index}
\end{equation}
\noindent and the TLI-g index from \citet{Piscarreta2024} is 
\begin{equation}
    \text{TLI-g}=\frac{F_{\lambda}(1.56 \mu m-1.58 \mu m)}{F_{\lambda}(1.625 \mu m-1.635 \mu m)}.
    \label{eq:tlig_index}
\end{equation}
The CHARIS wavelengths do not always line up with those at which these indices are evaluated.  In such cases, we evaluate the flux density at the closest available channel or compute averages between channels bracketing the prescribed index wavelength.

\citet{Piscarreta2024} notes field and intermediate-gravity objects but does not formally identify targets as ``low" or ``very low" gravity objects based on the \citeauthor{AllersLiu2013} conventions. We follow the convention that regions occupied exclusively by $<$30 Myr-old objects are scored as ``very low gravity" ($\delta$) and those containing only 0--300 Myr-old objects are classified as having ``low gravity" ($\gamma$). For classifications based on the H$_{\rm 2}$K index, we use \citet{Schneider2014}. Where there is an ambiguity in a classification, we list a range of scores\footnote{Given the sparse sampling of L/T transition objects with calculated indices in \citet{Piscarreta2024}, we emphasize that the exact gravity classification for each index may be uncertain. However, they provide a relative assessment of HIP 99770 b's gravity classification vs. other objects.}.  

Table \ref{tab:indices} summarizes the indices for HIP 99770 b, HR 8799 bc, HD 33632 Ab, and SIMPJ0956-1447, where the gravity score is shown in parentheses for each index and the average score and final classification is listed in the right column. The HR 8799 planets, especially HR 8799 b, show the strongest evidence for a low surface gravity.   Indices for SIMPJ0956-1447 and HD 33632 Ab are consistent with the objects' identifications as older field objects. HIP 99770 b's average gravity score is intermediate between those for the HR 8799 planets and field objects. Multiple $H$-band indices are consistent with a (very) low surface gravity object, while its H-cont index and both K-band indices are more similar to those for field dwarfs\footnote{\citet{AllersLiu2013} caution that the H continuum index is not always the most reliable indicator for gravity: HIP 99770 b and other targets analyzed here are on the edge of the spectral type range for which this index may be sensitive to gravity. Similarly, while \citet{Schneider2014} do estimate the H$_{2}$(K) indices for field L dwarfs vs. spectral type, the lack any measurements for low-gravity objects later than L8. Thus, it is also unclear whether the H$_{2}$(K) is a reliable diagnostic of gravity for objects like HIP 99770 b. Finally, the gravity classifications are sensitive to data quality. The K band spectral shapes for HR 8799 b from GRAVITY and OSIRIS show strong conflicts. Inspection of the GPI H band data for HR 8799 from \citet{Nasedkin2024} show that the detections of HR 8799 cd are weak/low significance per passband and the detection of HR 8799 e is just marginally above 5-$\sigma$ (5.8-$\sigma$) even in the wavelength-collapsed image: higher SNR detections in H band may lead to revisions in HR 8799 c's H band indices.}.

   \begin{figure*}
    \centering
       \vspace{-0.4in}
    \includegraphics[width=1\textwidth,trim=4mm 4mm 3mm 4mm,clip]{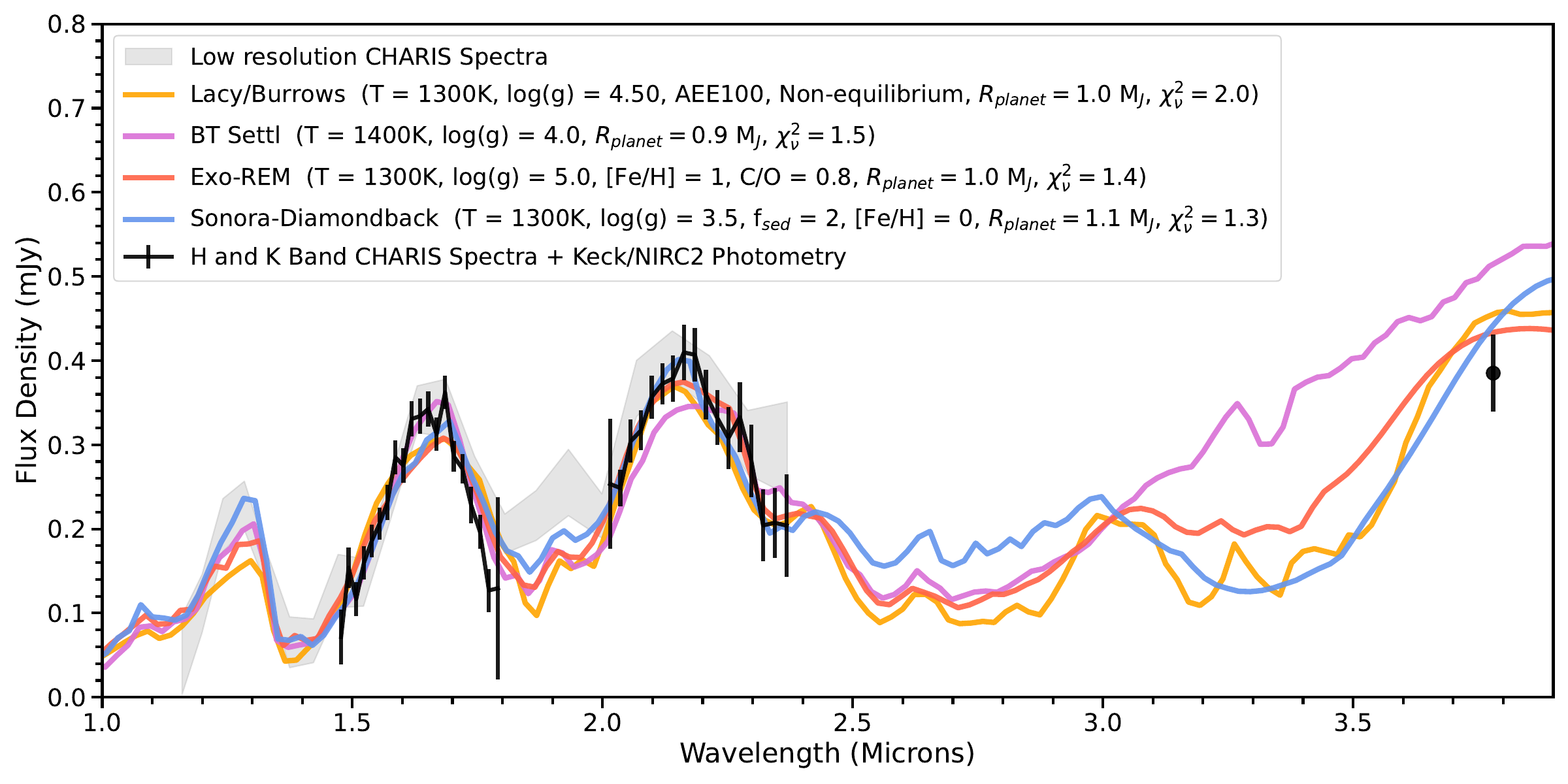}
    \vspace{-0.1in}
    \caption{The best fits from 4 atmospheric models, Lacy/Burrows, BT Settl, Exo-REM, and Sonora, are plotted against the H and K Band data as well as the photometry point in black (far right). The low resolution data is also plotted in gray for comparison at lower wavelengths but was not used for fitting the models.}
    \vspace{-0.in}
    \label{fig:atmod}
\end{figure*}

   \begin{figure*}
    \centering
    \includegraphics[width=0.45\textwidth,trim=0mm 0mm 0mm 0mm,clip]{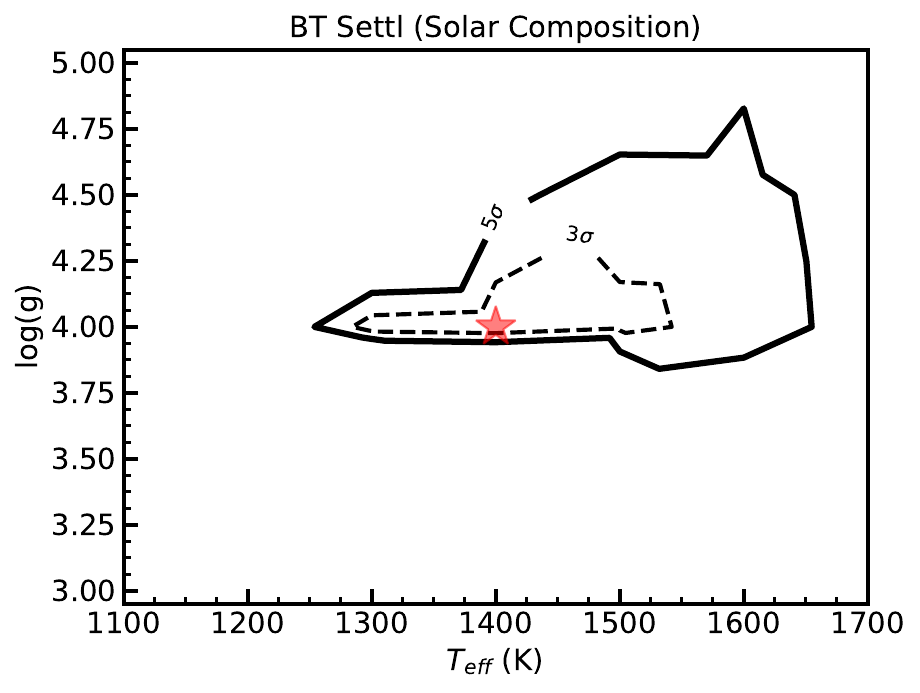}
        \includegraphics[width=0.45\textwidth,trim=0mm 0mm 0mm 0mm,clip]{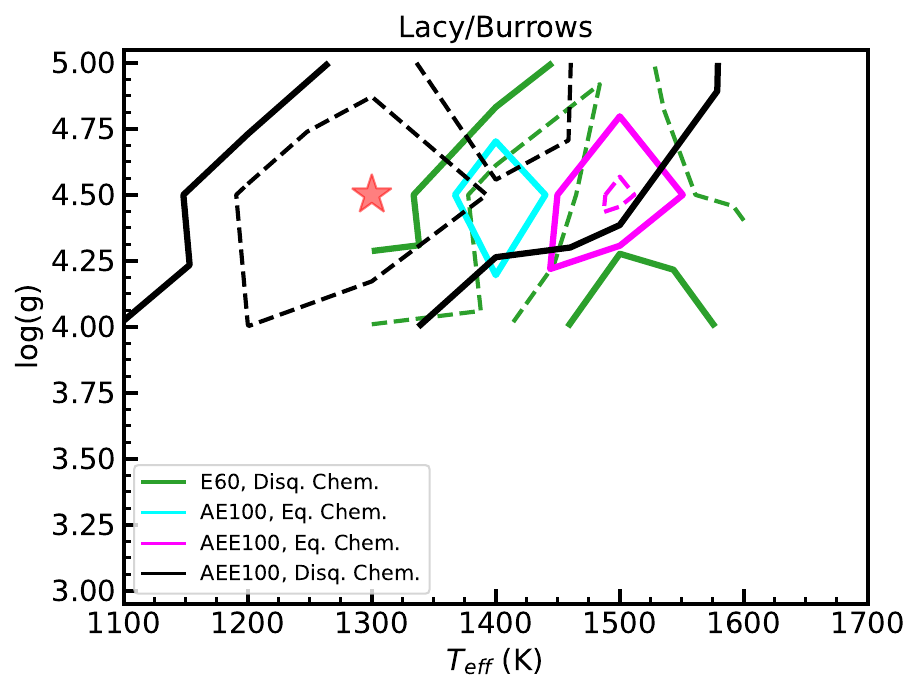}\\
     \includegraphics[width=0.45\textwidth,trim=0mm 0mm 0mm 0mm,clip]{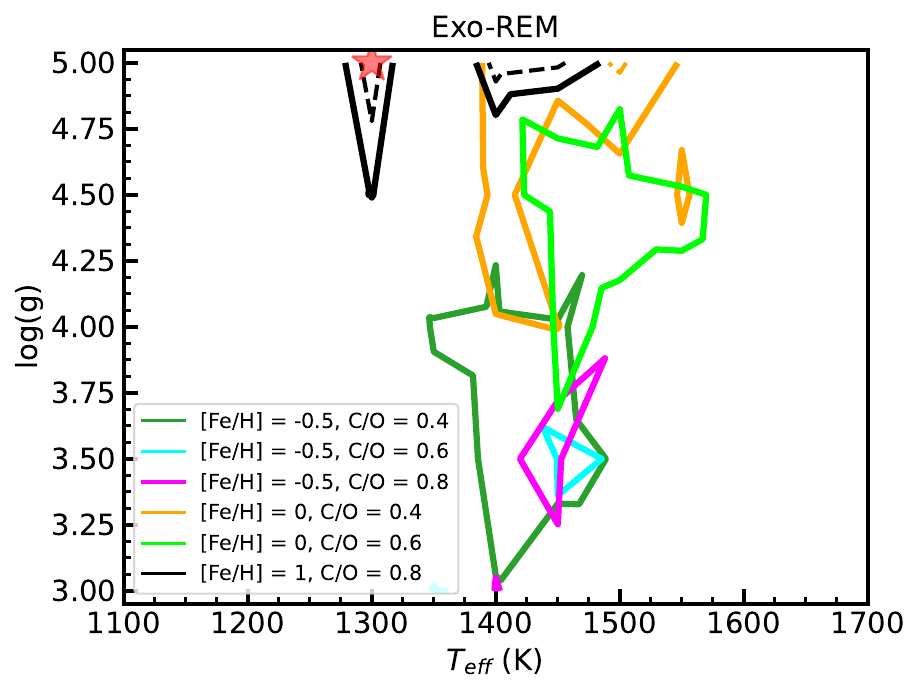}
     \includegraphics[width=0.45\textwidth,trim=0mm 0mm 0mm 0mm,clip]{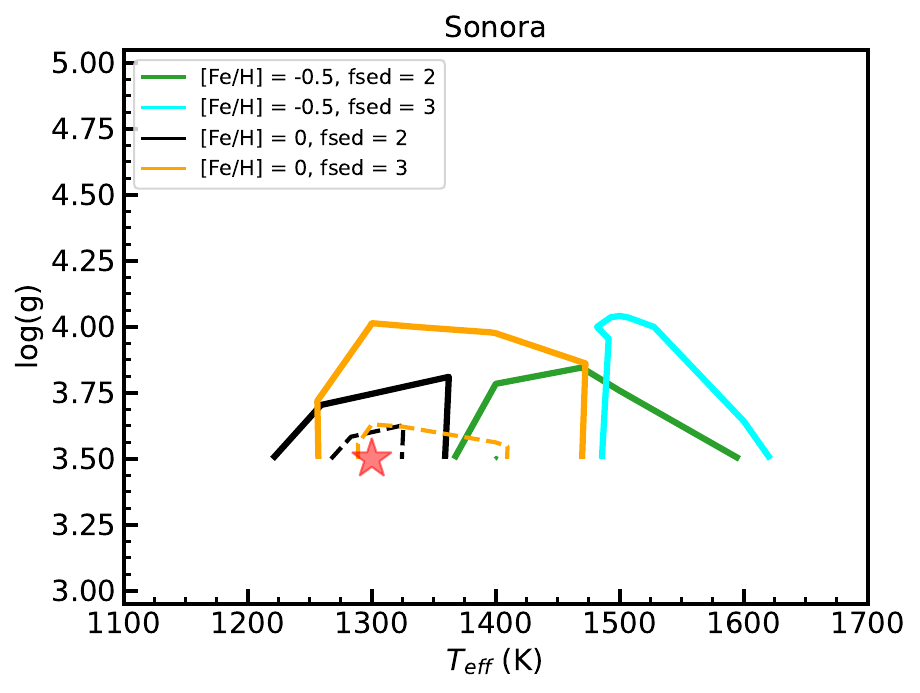}
    \vspace{-0.1in}
    \caption{The 3-$\sigma$ (dashed) and 5-$\sigma$ (solid) $\Delta \chi^{2}$ contours for the BT Settl, Lacy/Burrows, Exo-REM, and Sonora/Diamondback models in temperature/gravity space.  Black contours indicate the parameter combination yielding the best-fitting model (red star) from each grid.}
    \vspace{-0.in}
    \label{fig:atmodcontour}
\end{figure*} 

\subsection{Modeling HIP 99770 b's Atmosphere}

\subsubsection{Model Descriptions and Fitting Approach }
We also compared the CHARIS H and K spectra and Keck/NIRC2 photometry to several atmospheric models, including BT Settl \citep{Allard2012}, Exo-REM \citep{Charnay2018}, Sonora-Diamondback \citep{Morley2024}, and Lacy/Burrows \citep{Lacy2020}. These models vary temperature, gravity, chemistry, metallicity, and cloudiness. Table \ref{tab:atmmod} summarizes the parameter space explored for each grid: the grids are further described below.

\begin{itemize}
\item \textbf{BT Settl} 

BT Settl is a commonly used atmospheric model, adopting solar abundances from \citet{Asplund2009}, assuming equilibrium chemistry, and including a model for atmospheric dust and the formation of clouds that dissipate at the L/T transition. 
We used a subset of the full grid, shown in Table \ref{tab:atmmod}, $T_{eff}=400-2000K$, $log(g)=3.5-5$, and solar metallicity. 

\item \textbf{Lacy/Burrows}
The Lacy/Burrows model is based on \citet{Burrows2006} and was first used in \citet{Lacy2020}. It was most recently updated for \citet{Currie2023b} and used to fit the low resolution discovery of HIP 99770 b. Non-equilibrium carbon chemistry was added as well as updated molecular line lists and absorption cross sections. The clouds are parameterized into separate models AE, AEE, and E based on the clouds' structure and thickness at a given temperature and pressure. The atmospheric dust particle size also varies. This model spans a smaller temperature and gravity range than the other models discussed previously. We compare to the complete grid, which spans $T_{eff}=1100-1600K$, $log(g)=4-5$, and equilibrium or non-equilibrium carbon chemistry. 

\item \textbf{Exo-REM} 

Exo-REM is a more recent 1-dimensional radiative-equilibrium model, the newest version of which accounts for absorption and scattering of thermal radiation due to clouds \citep{Charnay2018}. It adopts abundances from multiple sources, including the ExoMol database \citep{Tennyson2012}. It allows for thermochemical equilibrium and some non-equilibrium chemistry. The model parameterizes the cloud particle size. They found the best results using simple microphysics which condenses silicate and iron clouds at the L-T transition and resulted in reddening for low-gravity objects as expected \citep{Charnay2018}. We compare to the entire grid, which spans $T_{eff}=400-2000K$, $log(g)=3-5$, $C/O=0.1-0.8$, and $[Fe/H]=0.32-10$.

\item \textbf{Sonora-Diamondback}

Sonora-Diamondback is the newest Sonora model \citep{Morley2024}. It includes refractory cloud species such as silicate clouds and assumes radiative-convective and chemical equilibrium. They parameterize clouds with \citet{Ackerman2001} cloud model. The vertical extent of the cloud is determined by the sedimentation efficiency ($f_{\rm sed}$), with small values corresponding to thick clouds and large values to thin compressed clouds. We used the complete grid which ranges $T_{\rm eff}$ = 900 -- 2400 $K$, log(g)= 3.5--5.5, $[M/H]$ = -0.5, 0.0, +0.5, and $f_{\rm sed}$ = 1, 2, 3, 4, 8, nc [no clouds].

\end{itemize}

 We find the best fit for the jth model again using $\chi^2$ statistics:
\begin{multline}
    \chi_{j}^{2} = R_{H, j}^{T}C_{H}^{-1}R_{H,j} + R_{K, j}^{T}C_{K}^{-1}R_{K,j}\\
    + \sum_{i}(f_{phot,i}-\alpha_{j}~F_{phot,ij})^{2}/\sigma_{phot,i}^{2},
    \label{eq:chi2}
\end{multline}
where $R_{H,j}$ and $R_{K,j}$ equal  $f_{spec} -\alpha_{j} F_{spec}$ the difference between the measured (f) and model-predicted (F) flux density for each passband, $C_{H}$ and $C_{K}$ are the covariance for each passband, $f_{phot,i}$ is the measured photometry, $F_{phot,ij}$ is the model’s predicted photometry, $\sigma_{phot,i}$ is the photometric uncertainty, and $\alpha_{j}=(R/D)^2$ is scaled for each model to minimize the $\chi^2$. We assume a distance of 40.74 parsecs to the system \citep{GAIA2018}.

\subsubsection{Results} 
Figure \ref{fig:atmod} compares the HIP 99770 b H and K band spectra and NIRC2 photometry point (all shown in black) to best-fitting atmospheric models from each grid.  The Sonora Diamondback grid model with $T_{\rm eff}$ = 1300 $K$, log(g) = 3.5, solar metallicity, and $f_{\rm sed}$ = 2 and a radius of 1.056 Jupiter radii quantitatively provides the best fit to the HIP 99770 b data ($\chi_{\nu}^{2}$ = 1.33). Other grids produce models that generally reproduce the peak and wings of the $H$ and $K$ band spectra: their higher $\chi^{2}$ values reflect the high precision of the data coupled with the sparse sampling of the grid in key parameter space. 

The grids show broad agreement on the best-fitting temperature for HIP 99770 b (1300-1400 $K$). The surface gravity is far more poorly constrained although there is a hint of a preference for values at or below log(g) = 4.5. The best-fitting radii are systematically smaller than values from the Sonora evolutionary models for 10--20 $M_{\rm Jup}$, 100--200 Myr old objects, though grids other than BT-Settl yield values within 20-25\% of predictions.

Figure \ref{fig:atmodcontour} further explores the fit of each grid in temperature vs. gravity. While the BT-Settl models assume solar metallicity and equilibrium chemistry, others include additional varied parameters (e.g. C/O ratio for the Exo-REM models or $f_{\rm sed}$ for Sonora Diamondback). We plot representative $\Delta \chi^{2}$ contours for each of these assumptions separately. The BT Settl models have a relatively large 5-$\sigma$ $\Delta \chi^{2}$ surface, especially in temperature. The Lacy/Burrows models show some dependence of the preferred temperature (but not surface gravity) on the adopted cloud model and chemistry.  

The best-fitting Sonora models lie at the edge of the available grid in surface gravity: all favor low values regardless of metallicity. In contrast, for Exo-REM there appears to be correlation between the preferred temperature/gravity parameter space and metallicity, where low-metallicity models favor low-gravity solutions. The best-fitting Exo-REM model lies at an extreme value in both metallicity (ten times solar) and carbon-to-oxygen ratio (0.8). It appears to represent a small island of best-fitting solutions from this grid in an otherwise poorly fitting region: models with [Fe/H] = 1 and lower C/O or C/O = 0.8 and lower metallicity (not shown) do not reproduce HIP 99770 b's spectrum.  Similarly, Sonora models with extremely thick clouds ($f_{\rm sed}$ = 1), far thinner clouds ($f_{\rm sed}$ = 8), or no clouds at all (not shown) do not reproduce HIP 99770 b's spectrum.

\begin{deluxetable*}{llllllll}
     \tablewidth{0pt}
     \tabletypesize{\scriptsize}
    \tablecaption{Atmosphere Model Parameter Space\label{tab:atmmod}}
    \tablehead{\colhead{Model} & \colhead{$T_{\rm eff} (K)$} &  \colhead{log(g)} &\colhead{C/O} & \colhead{[M/H]} & \colhead{$f_{\rm sed}$} & \colhead{Best Fit} & \colhead{$\chi_{\rm \nu}^2$}}
    \startdata
   BT Settl & 400-2000 &  3.5-5 & Solar & Solar & N/A & $T_{\rm eff} = 1400 K$, $log(g) = 4$, $R =0.892 R_{\rm J}$ & 1.476\\
    Exo-REM & 400-2000 &  3-5 & 0.1-0.8 & 0.32-10 & N/A & $T_{\rm eff} = 1300 K$, $log(g) = 5$, $R =1.015 R_{\rm J}$ & 1.441\\
    Sonora & 900-2400  & 3.5-5.5  &  Solar & 0.32, 0, 3.2 & 1, 2, 3, 4, 8, nc & $T_{\rm eff} = 1300 K$, $log(g) = 3.5$, $R =1.056 R_{\rm J}$ & 1.33\\
    Lacy/Burrows & 1100-1600  &  4-5  & See Note &  Solar & N/A  & $T_{\rm eff} = 1300 K$, $log(g) = 4.5$, $R =1.003 R_{\rm J}$ & 2.016\\
    \enddata
    \tablecomments{Lacy/Burrows carbon chemistry: equilibrium chemistry and non-equilibrium/0.1×CH4}

    \end{deluxetable*}

 \begin{figure*}
\centering
   \includegraphics[width=0.45\textwidth,trim=5mm 10mm 4mm 20mm,clip]{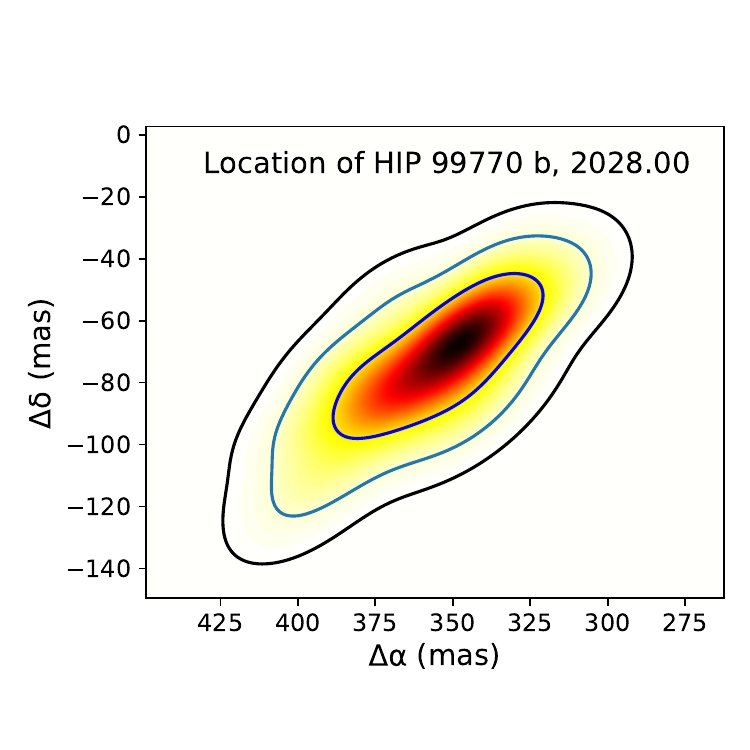}
   \includegraphics[width=0.5\textwidth,trim=10mm 6mm 7mm 10mm,clip]{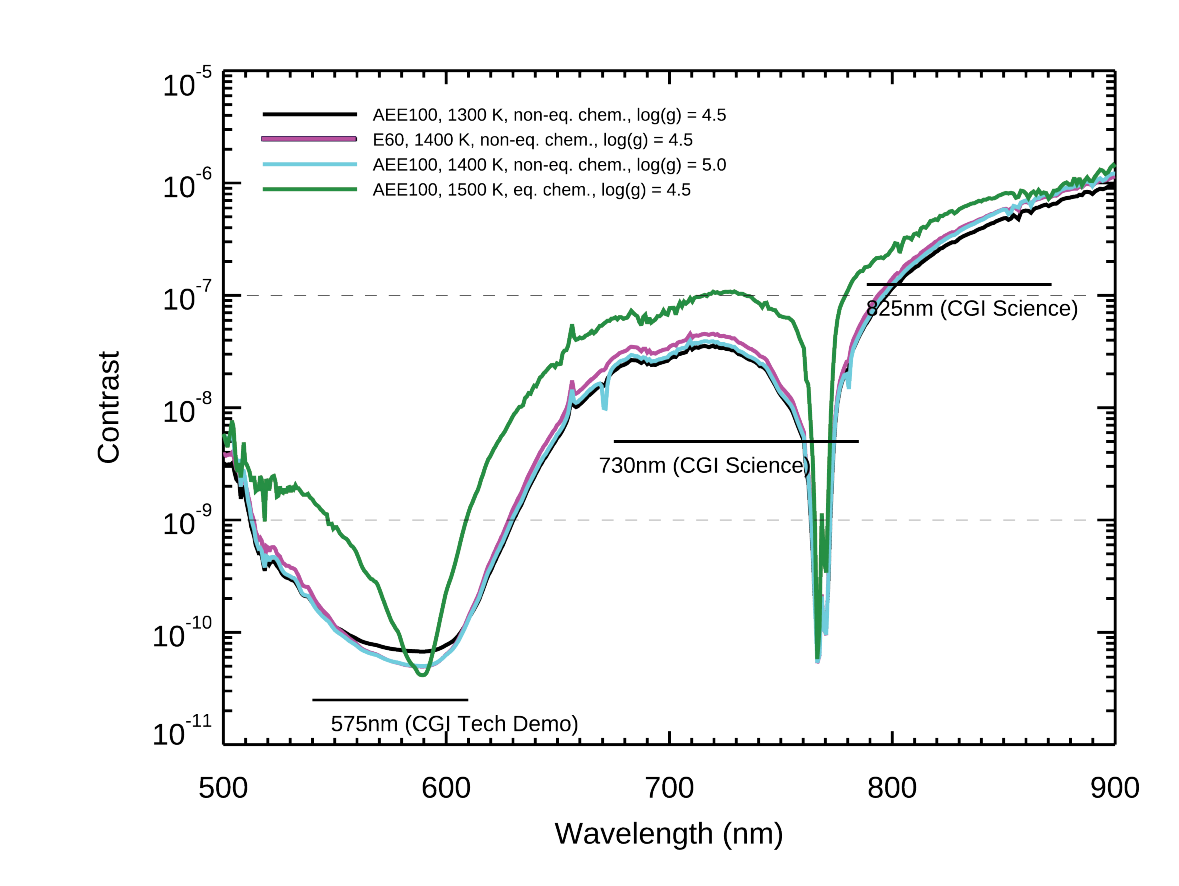}
   \caption{(Left) Predicted location of HIP 99770 b from the best-fitting \texttt{orvara} orbits (assuming flat priors on the companion mass) in January 2028.   (Right) Predicted contrast vs. wavelength for HIP 99770 b from the best-fitting Lacy/Burrows models considered in this paper. }
   \label{fig:location} 
\end{figure*}

 \subsection{Predictions for Detectability with Roman CGI at Optical Wavelengths}
As HIP 99770 b is a self-luminous companion imaged at $\approx$0\farcs{}4 orbiting a very bright star, we consider its suitability for detection and characterization with the Nancy Grace Roman Telescope's Coronagraph Instrument (CGI) during the instrument's Technology Demonstration \citep[the ``tech demo"][]{Kasdin2020}. At 575 nm, the instrument should generate a dark hole between 0\farcs{}15 and 0\farcs{}45. Roman CGI's main tech demo requirement is to demonstrate $<$10$^{-7}$ on a V $\lesssim$5 star at the 575 nm bandpass within 6--9 $\lambda$/D from the star (0\farcs{}3--0\farcs{}45) in the instrument's dark hole region. Tech demo time subsequent to a successful achievement of this goal may be spent on photometric, astrometric, and spectroscopic characterization of self-luminous planets in thermal emission and mature planets in reflected light.

To evaluate HIP 99770 b’s suitability as a CGI target, we first determine whether it will be located within the CGI dark hole during the first few years of the Roman mission during which the tech demo will likely be carried out. As a fiducial estimate, we used \texttt{orvara} dynamical modeling to predict the planet’s position in January 2028, early in the mission and bracketing the likely earliest opportunities for targeting in Fall 2027 and Spring 2028. As shown in Figure \ref{fig:location} (left), HIP 99770 b is predicted to lie at an angular separation of $\approx$0\farcs{}35 during the first year during which it could be targeted.  This is well within the CGI dark hole at 575 nm and inside the 0\farcs{}45 working angle for tech demo observations.

The Lacy/Burrows models allow for a robust estimate of optical contrasts \citep{Lacy2020}. To model the planet's expected contrast in the Roman CGI passbands, we consider the top four best-fitting Lacy/Burrows atmospheric models (see Section 4.3) \citep{Currie2023b} and compute a contrast ratio drawn from the MILES library spectrum of the HIP 99770 A primary \citep{Sanchez-Blazquez2006}. Figure \ref{fig:location} (right) shows the predicted contrasts in different passbands. At 575 nm, the contrast values are 7.8 $\times$10$^{-11}$ to 4.4$\times$10$^{-10}$. These are too faint for CGI to feasibly yield a detection given current performance estimates.

However, contrasts significantly improve at longer wavelengths, 2.9--8.1$\times$10$^{-8}$ at the 730 nm passband for low-resolution spectroscopy and 3.6--5.7$\times$10$^{-7}$ for wide-field imaging at 825 nm. For these two passbands, the expected CGI dark hole regions are $\approx$0\farcs{}18--0\farcs{}55 and $\approx$0\farcs{}45--1\farcs{}4. HIP 99770 b thus should be well suited for spectroscopic follow up. While the companion lies slightly interior to the nominal dark hole region at 825 nm, the contrast is sufficiently modest enough that the companion may be detectable depending on CGI's in-flight performance.  Because of its detectability at 730 nm, HIP 99770 b is now proposed in the Roman CGI white paper call as an early “spectroscopic technological demonstration" focused on spectral extraction and atmospheric parameter estimation over a range of contrasts and contamination from exozodi/debris disk emission \citep{Currie2025}.

 \section{Discussion}

Our higher-resolution (R $\sim$ 70) H and K band integral field spectroscopy of HIP 99770 b with SCExAO/CHARIS modeled jointly with Hipparcos and Gaia supported the findings from the previous observations \citep{Currie2023b} but with slightly improved constraints. Incorporating a relative radial velocity measurement from \citet{Zhang2024} refines the dynamical mass estimate to 15.0$_{-4.4}^{+4.5}$ $M_{\rm Jup}$ with a flat prior and 13.1$_{-5.2}^{+4.8}$ $M_{\rm Jup}$ with a log-uniform prior. The semimajor axis we find is slightly smaller (15.8$_{-1.0}^{+3.1}$au for flat prior) than what was reported in the discovery paper, while the eccentricity is slightly larger (0.29$_{-0.15}^{+0.12}$ for flat prior), disfavoring a circular orbit.   We find consistent results with and without incorporating recent VLTI/GRAVTIY astrometry, although the inclusion of the latter yields a slightly smaller median posterior value for the eccentricity with a narrower 68\% confidence interval.

Atmospheric modeling favors a temperature near 1300 K and surface gravity in the range log(g) $\approx$ 3.5–4.5, though with significantly more scatter in the gravity. 
Empirical comparisons suggest that HIP 99770 b is a L6--L9.5 dwarf with a surface gravity intermediate between that of the HR 8799 planets and older field brown dwarfs of the same temperatures but carbon that may be in chemical equilibrium.  Our results suggest that HIP 99770 b may provide a window into the atmospheric evolution of substellar objects at the L/T transition and a valuable context for testing atmospheric models of cloud evolution, chemical mixing, and planet formation. 

Following \citet{ElMorsy2024}, our study reinforces the value of combining RV measurements extracted from high-resolution spectroscopy in dynamical modeling of planets and brown dwarfs.  Similar data for other stars with very modest astrometric accelerations like HIP 99770's will improve dynamical mass constraints for their companions.   Spectroscopy at higher resolutions complements probes of planet and brown dwarf atmospheres with lower resolution spectra from CHARIS and similar instruments \citep[e.g.][]{Konopacky2013,Zhang2024}.   The Gaia DR4 release expected in 2026 should improve astrometric measurements for HIP 99770 to more precisely constrain the mass of its companion.

Although likely inaccessible to Roman CGI at 575 nm for the tech demo for most reasonable assumptions about its atmosphere, HIP 99770 b is a promising candidate for spectroscopic characterization at 730 nm and imaging at 825 nm during the tech demo phase of the mission and beyond, offering a path toward deeper insights into its metallicity and atmospheric chemistry. Optical follow-up at these wavelengths could reveal absorption features such as potassium and sodium which would indicate a more metal-rich atmosphere and the broad continuum shape at these higher wavelengths could be indicators for the dynamics and structure of the atmosphere \citep{Lacy2020}.

Revisiting HIP 99770 b with JWST could also provide an opportunity to probe its atmosphere in greater detail, particularly at wavelengths where a detection is difficult to achieve from the ground. JWST’s angular resolution and sensitivity at 3–5 $\mu$m would allow the presence of CO, CO$_2$, and CH$_4$ to be measured, offering further insights into the planet’s chemical composition, metallicity, and cloud structure \citep{Balmer2025, Franson2024}. Stronger CO$_2$ absorption would suggest a more metal rich atmosphere. Similar studies have recently been done for AF Lep b \citep{Franson2024}, 51 Eri b,  and the HR 8799 planets \citep{Balmer2025}. Combined with similar data for other objects, JWST follow-up of HIP 99770 b would help advance our understanding of L/T transition and intermediate-gravity objects and further help to place it into the context of the broader population of imaged planets.

\begin{acknowledgments}
We thank the anonymous referee for helpful comments that improved the scientific quality and presentation of this work.

\indent The authors wish to acknowledge the very significant cultural role and reverence that the summit of Mauna Kea holds within the Hawaiian community.  We are most fortunate to have the opportunity to conduct observations from this mountain.

\indent The development of SCExAO was supported by JSPS (Grant-in-Aid for Research \#23340051, \#26220704 \& \#23103002), Astrobiology Center of NINS, Japan, the Mt Cuba Foundation, and the director's contingency fund at Subaru Telescope.  CHARIS was developed under the support by the Grant-in-Aid for Scientific Research on Innovative Areas \#2302.

\indent T.C. is supported by National Science Foundation Astronomy and Astrophysics Grant \#2408647.\\

\end{acknowledgments}

\bibliography{bibliography}{}
\bibliographystyle{aasjournal}

\appendix
\renewcommand\thefigure{\thesection.\arabic{figure}}    
\section{Additional Figures}
\setcounter{figure}{0} 

\begin{figure*}[h]
    \begin{flushright}
    \includegraphics[width=1\textwidth]{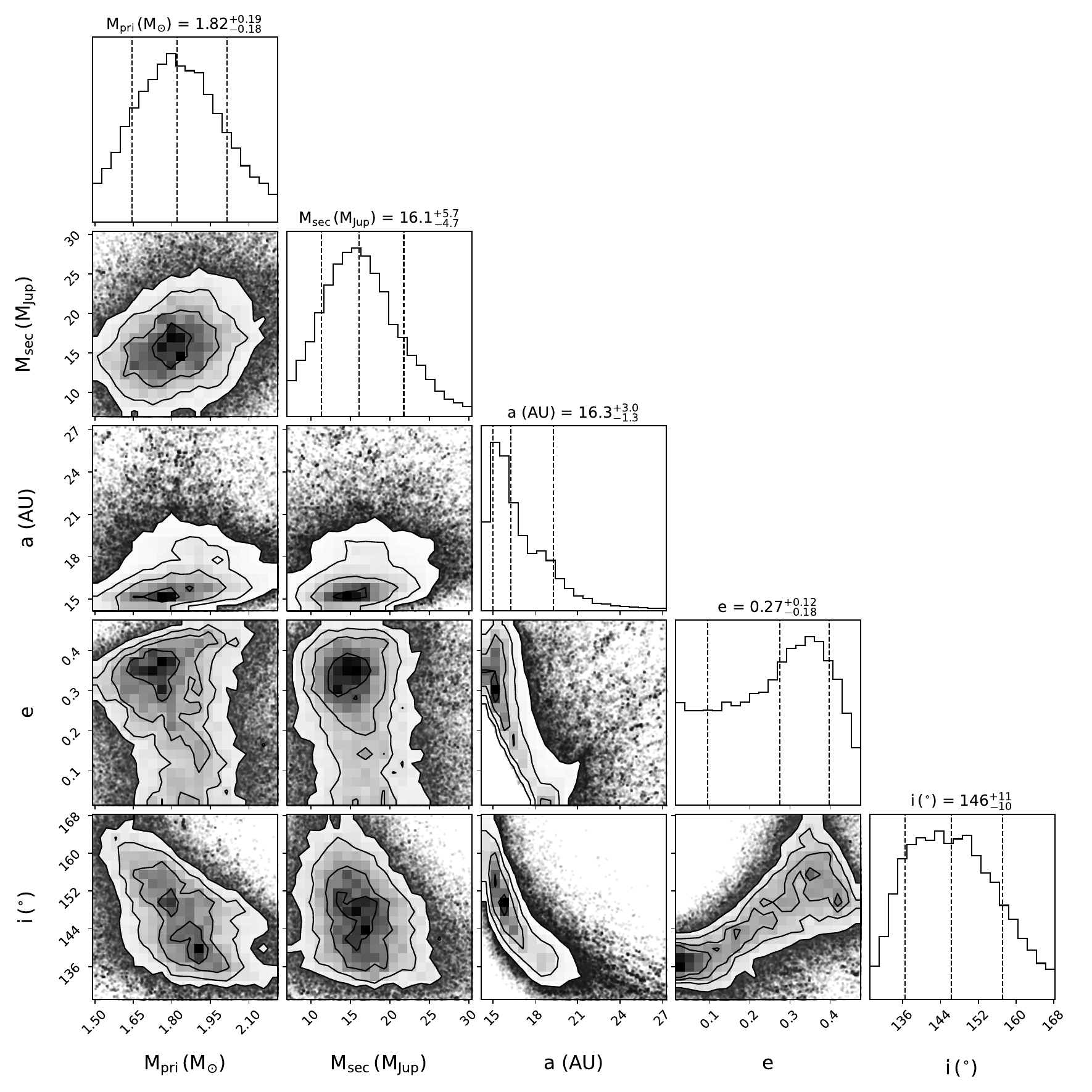} \\
    \vspace{-0.02\textwidth}
    \end{flushright}
    \caption{Corner plot showing MCMC posterior distributions for the flat prior without RV measurements from KPIC. Fit using only absolute astrometry from HGCA and relative astrometry (from both initial discovery and follow-up). The contours show the $68\%$ $(1\sigma)$, $95\%$ $(2\sigma)$, and $99\%$ $(3\sigma)$ confidence intervals.}
    \vspace{-0.in}
    \label{fig:corner_norv}
\end{figure*}

\begin{figure*}
   \includegraphics[width=0.34\textwidth,trim=0mm 3mm 2mm 0mm,clip]{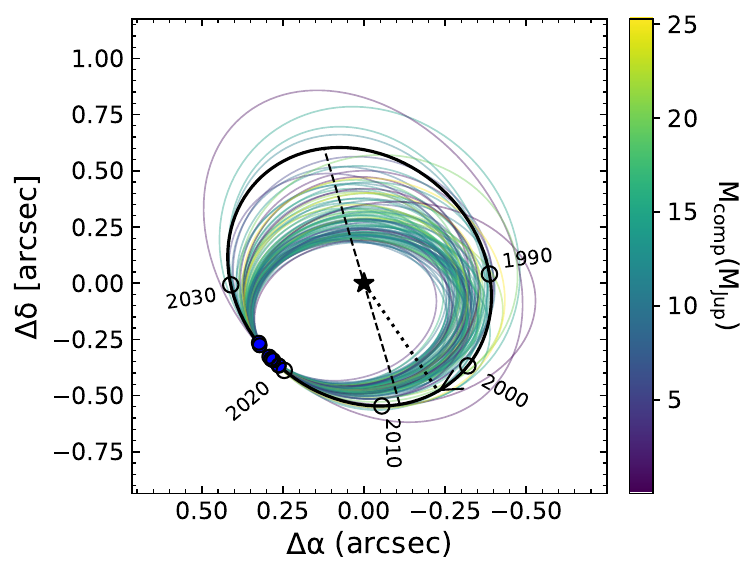}
    \includegraphics[width=0.32\textwidth,trim=0mm 3mm 2mm 0mm,clip]{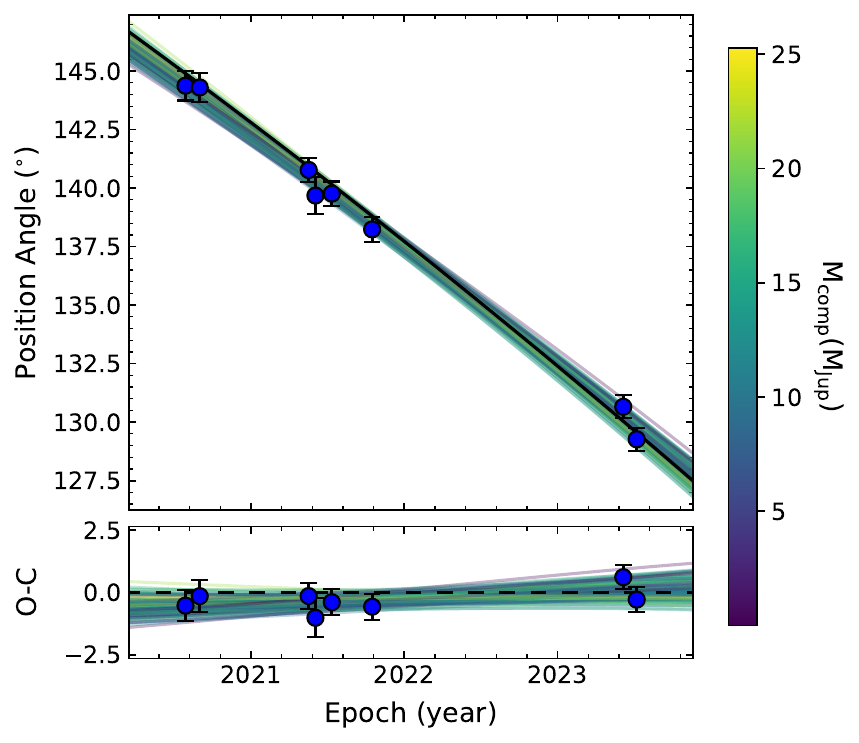}
    \includegraphics[width=0.32\textwidth,trim=0mm 3mm 2mm 0mm,clip]{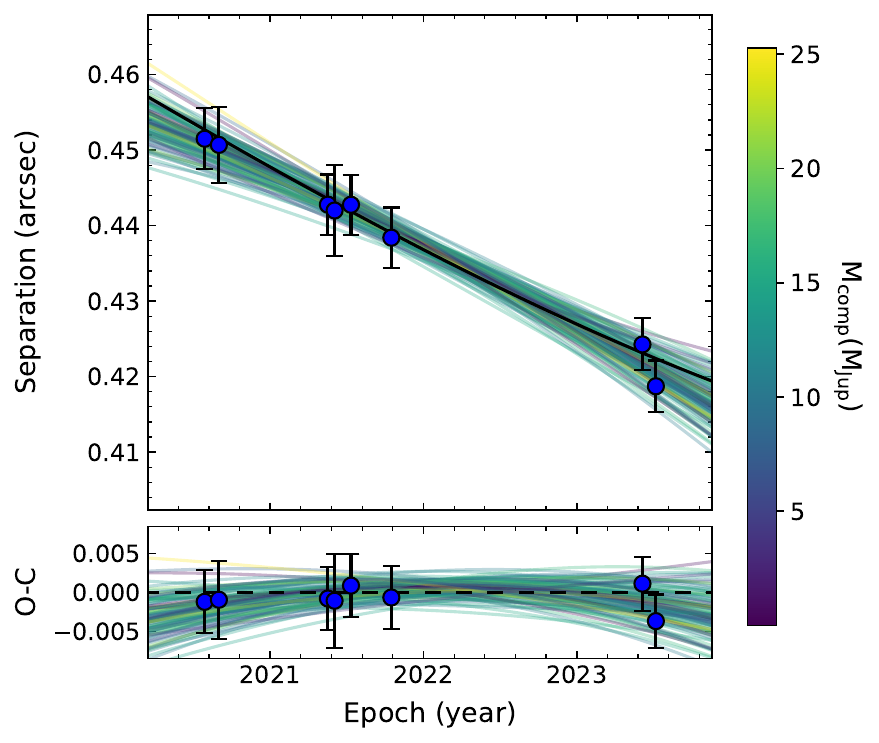}\\
     \vspace{-0.15in}
    \caption{Log-uniform orbit fitting with \texttt{orvara}: (Left) Predicted orbits with most likely orbit in black along with 100 randomly selected orbits (color coded by mass) from the MCMC posterior distribution. Blue circles show our data, while empty circles show predicted location per epoch. (Center) Position angle vs epoch (Right) Separation vs epoch}
    \label{fig: orbit_lognorm}
\end{figure*}

\begin{figure*}
   \includegraphics[width=0.34\textwidth,trim=0mm 3mm 2mm 0mm,clip]{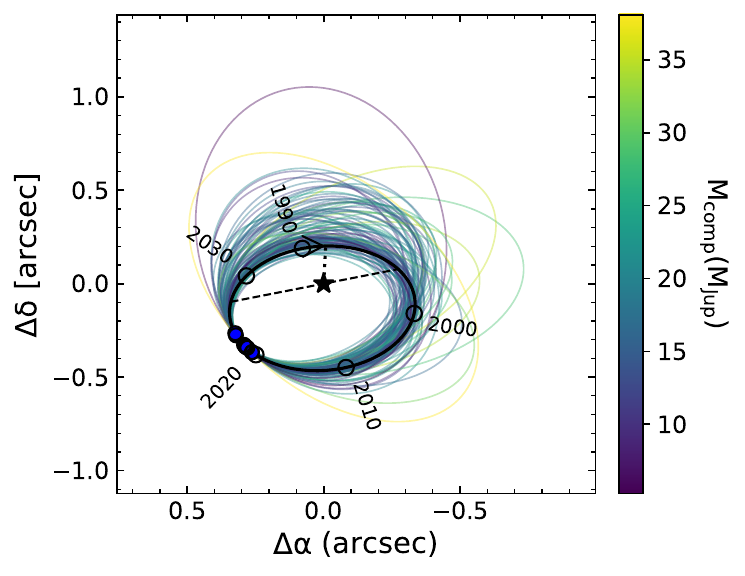}
    \includegraphics[width=0.32\textwidth,trim=0mm 3mm 2mm 0mm,clip]{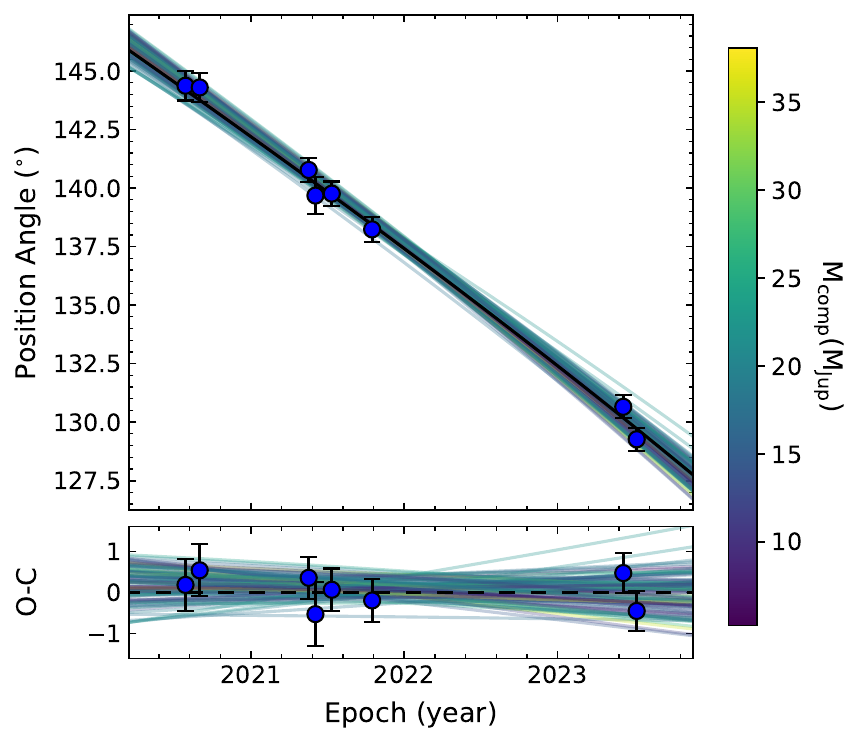}
    \includegraphics[width=0.32\textwidth,trim=0mm 3mm 2mm 0mm,clip]{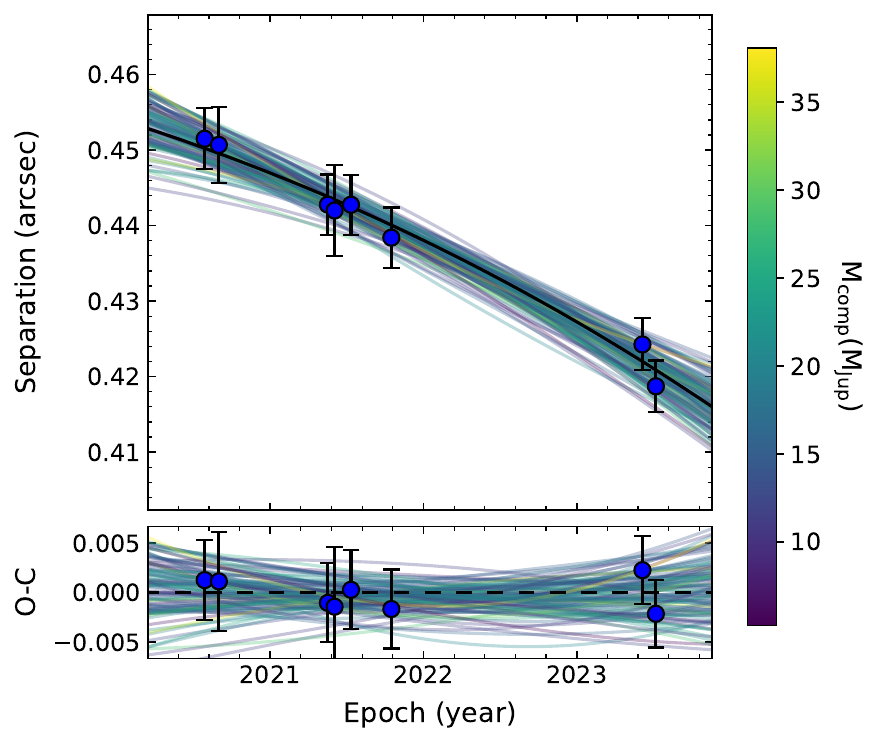}\\
     \vspace{-0.15in}
    \caption{No RV orbit fitting with \texttt{orvara}: (Left) Predicted orbits with most likely orbit in black along with 100 randomly selected orbits (color coded by mass) from the MCMC posterior distribution. Blue circles show our data, while empty circles show predicted location per epoch. (Center) Position angle vs epoch (Right) Separation vs epoch}
    \label{fig: orbit_norv}
\end{figure*}

\begin{figure*}
   \includegraphics[width=1\textwidth,trim=0mm 3mm 2mm 0mm,clip]{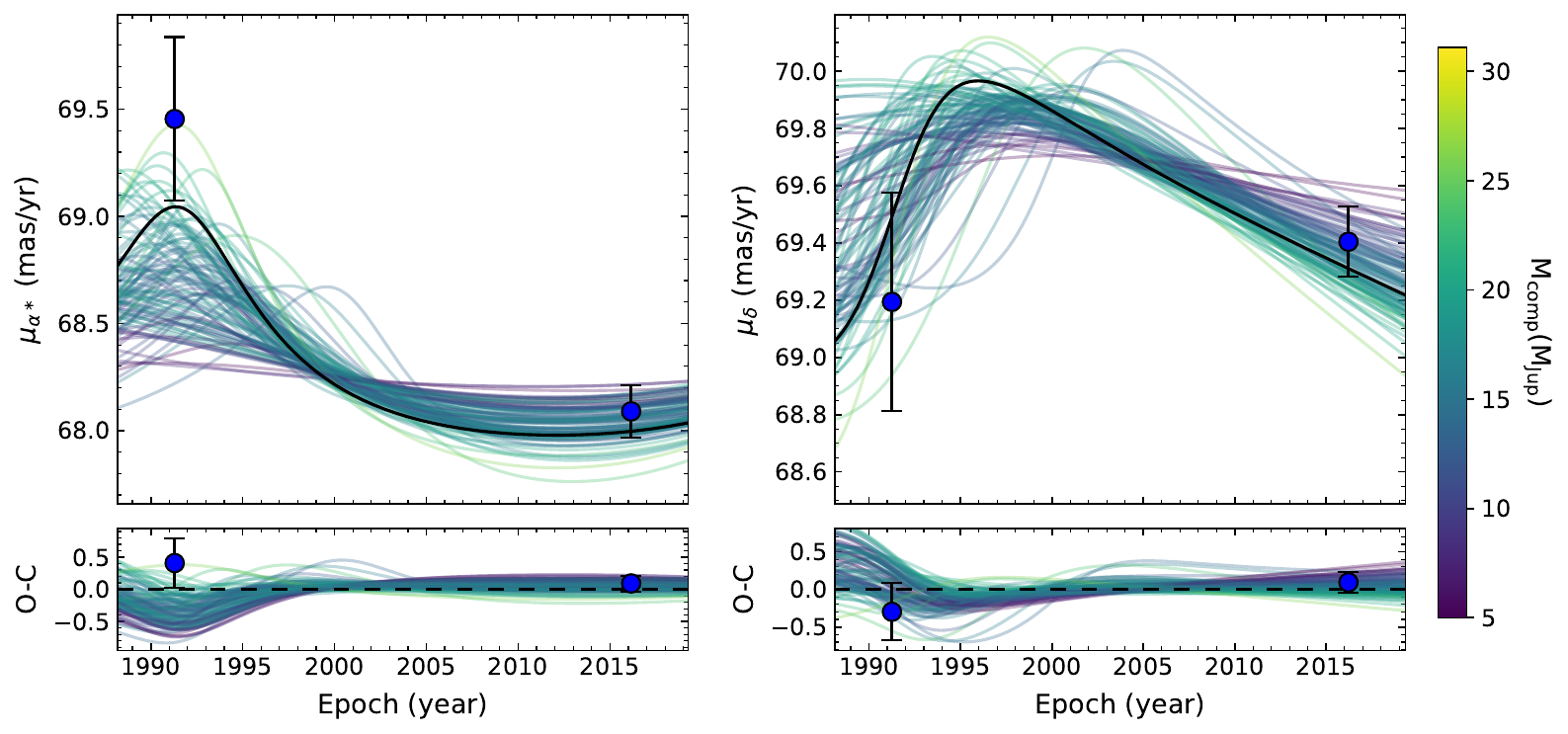}\\
     \vspace{-0.15in}
    \caption{The predicted astrometric motion in right ascension (Left) and declination (Right) for the flat prior. Shown are 100 randomly selected orbits (color coded by mass) from the MCMC posterior distribution and our data with blue circles.}
    \label{fig: pm_fp}
\end{figure*}

\begin{figure*}
   \includegraphics[width=1\textwidth,trim=0mm 3mm 2mm 0mm,clip]{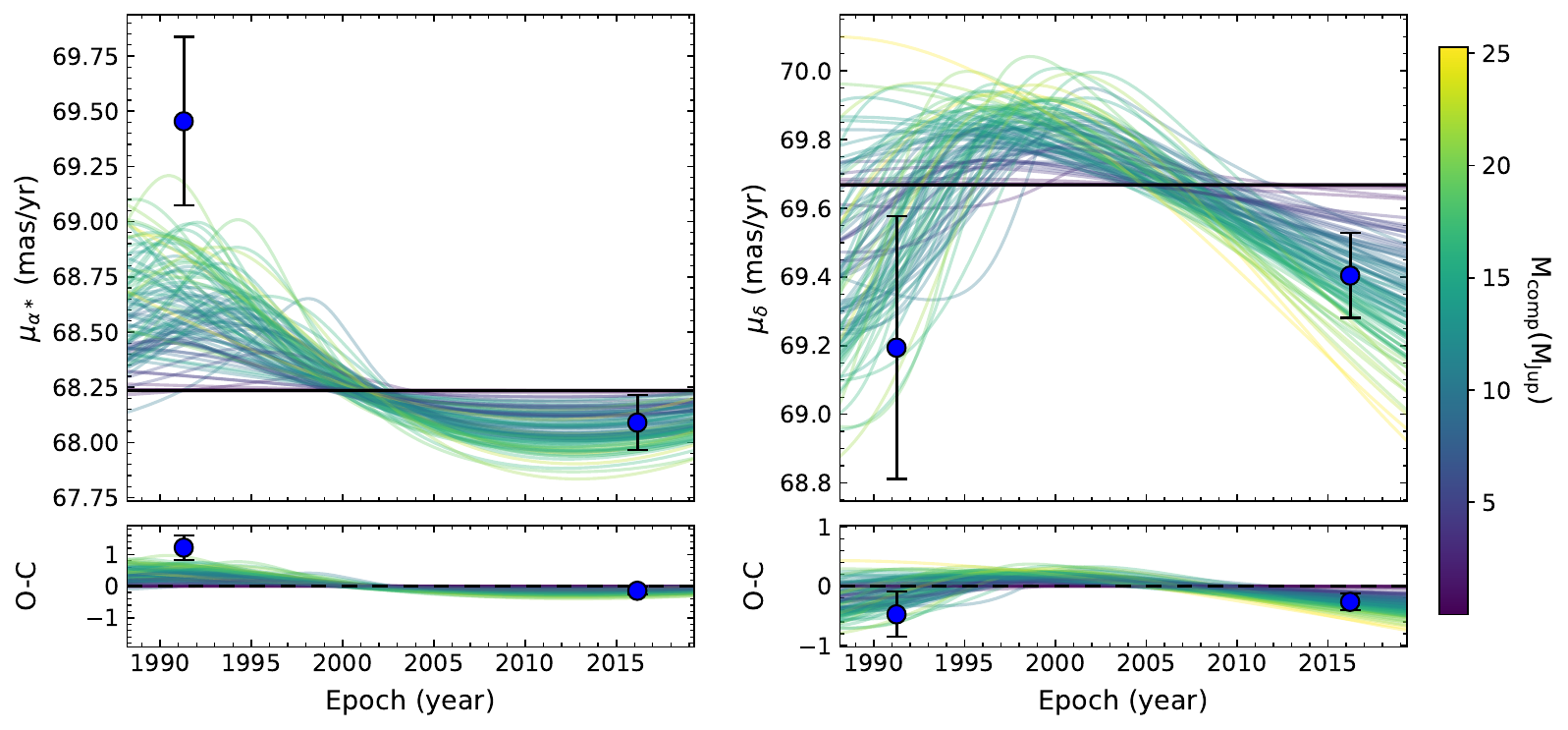}\\
     \vspace{-0.15in}
    \caption{The predicted astrometric motion in right ascension (Left) and declination (Right) for the log-uniform prior. Shown are 100 randomly selected orbits (color coded by mass) from the MCMC posterior distribution and our data with blue circles.}
    \label{fig: pm_ln}
\end{figure*}

\begin{figure*}
   \includegraphics[width=1\textwidth,trim=0mm 3mm 2mm 0mm,clip]{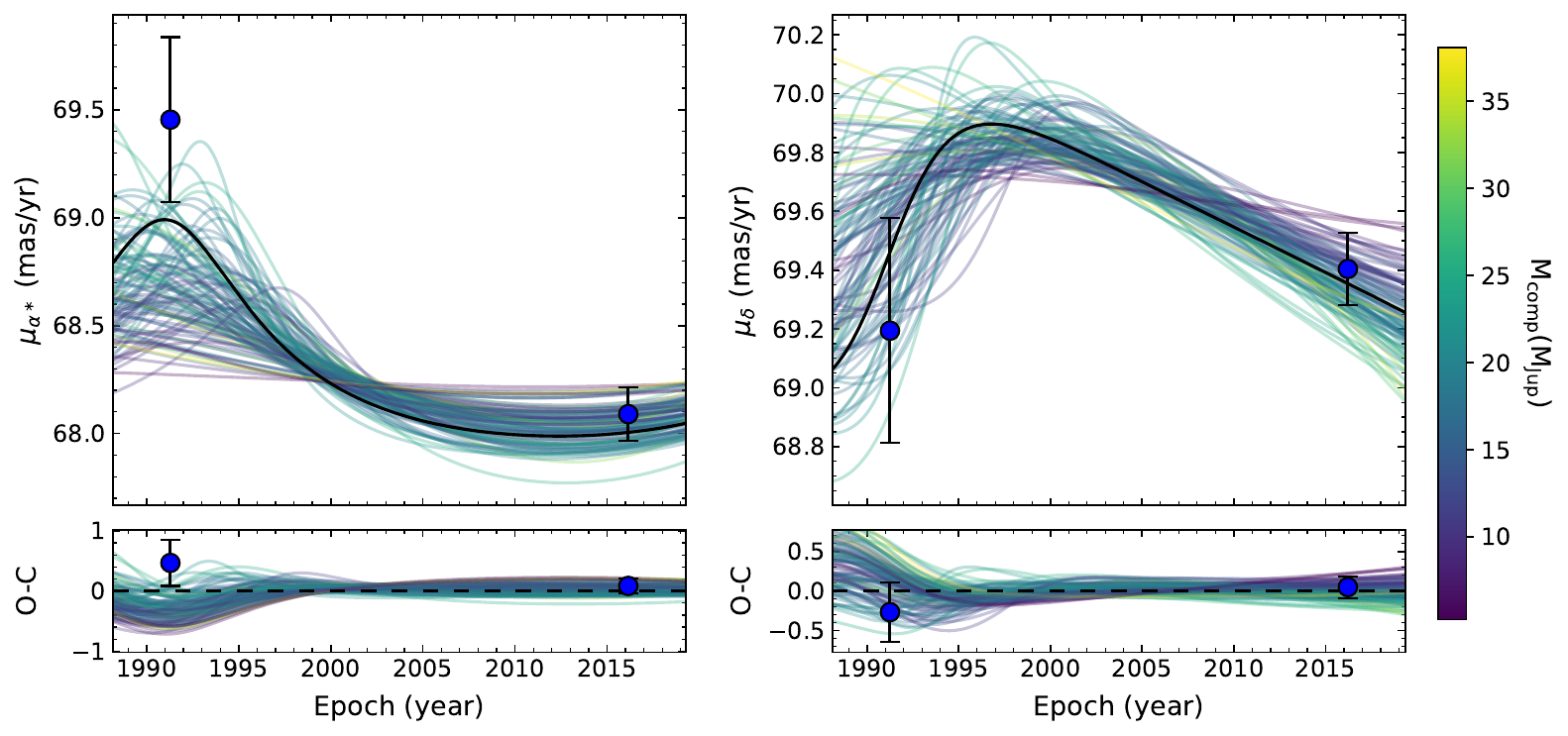}\\
     \vspace{-0.15in}
    \caption{The predicted astrometric motion in right ascension (Left) and declination (Right) for the the flat prior without RV data. Shown are 100 randomly selected orbits (color coded by mass) from the MCMC posterior distribution and our data with blue circles.}
    \label{fig: pm_norv}
\end{figure*}

\begin{figure*}
    \includegraphics[width=1\textwidth,trim=0mm 3mm 2mm 0mm,clip]{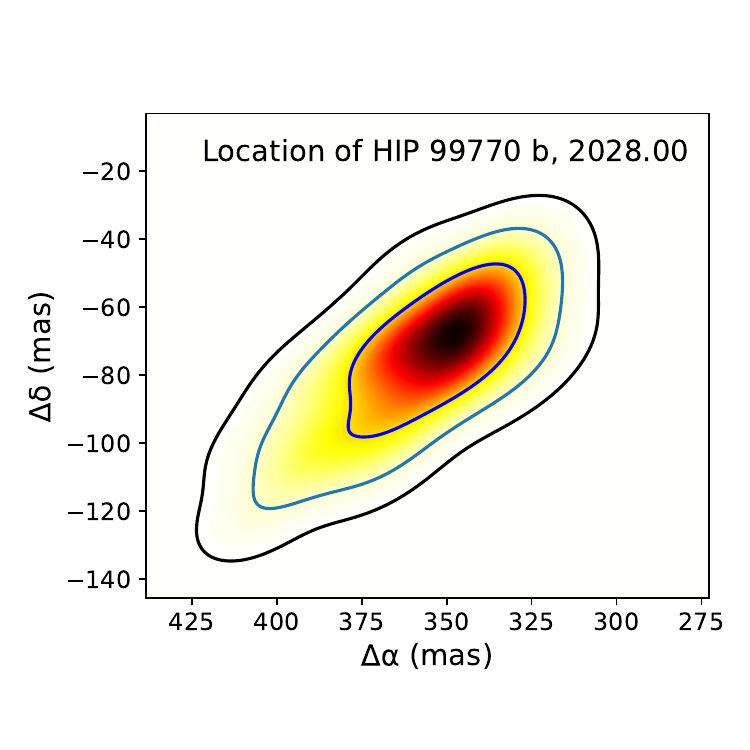}\\
     \vspace{-0.15in}
    \caption{Predicted location of HIP 99770 b from the best-fitting \texttt{orvara} orbits (assuming log-uniform priors on the companion mass) in January 2028.}
    \label{fig: predict_ln}
\end{figure*}

\begin{figure*}
    \includegraphics[width=1\textwidth,trim=0mm 3mm 2mm 0mm,clip]{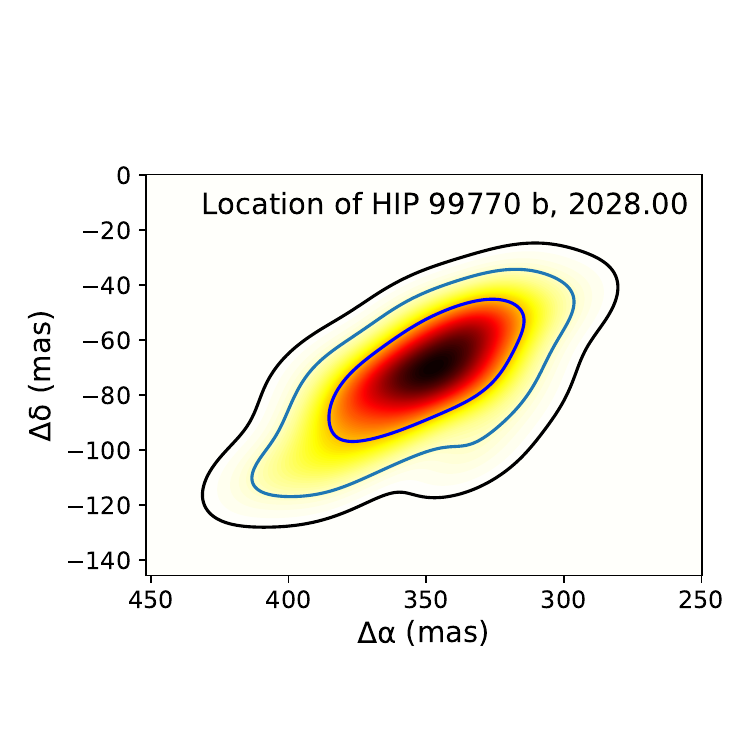}\\
     \vspace{-0.15in}
    \caption{Predicted location of HIP 99770 b without RV data from the best-fitting \texttt{orvara} orbits (assuming flat priors on the companion mass) in January 2028.}
    \label{fig: predict_norv}
\end{figure*}

Figure \ref{fig:corner_norv} shows the corner plot for the flat prior without RV data from KPIC. Figures \ref{fig: orbit_lognorm} and \ref{fig: orbit_norv} show predicted orbits (Left), position angle vs epoch (Center), and separation vs epoch (Right) for the log-uniform and no RV data flat prior respectively.
Figures \ref{fig: pm_fp}, \ref{fig: pm_ln}, and \ref{fig: pm_norv} show the predicted astrometric motion in right ascension (Left) and declination (Right) for the flat prior, log-uniform prior, and the flat prior without RV data respectively.
Figures \ref{fig: predict_ln} and \ref{fig: predict_norv} show the predicted location of HIP 99770 b in 2028 for the log-uniform prior and the flat prior without RV data.

\end{document}